# Hydrogen-induced volume expansion in hexagonal close-packed iron: Effects of pressure and temperature

Yuichiro Mori[1,2], Masahiro Takano[1], Hiroyuki Kagi[1], Katsutoshi Aoki[1], Sho Kakizawa[3], Noriyoshi Tsujino[3], Yuji Higo[3]

[1] Graduate School of Science, The University of Tokyo, Bunkyo, Tokyo 113-0033, Japan

[2] Department of Earth and Planetary Sciences, Yale University, New Haven, CT 06511, USA

[3] Japan Synchrotron Radiation Research institute, Sayo, Hyogo, 679-5198, Japan

## Key points

- We experimentally determined *PVT* relation of hcp-$FeH_x$.
- Hydrogen-induced volume expansion of hcp-$FeH_x$ has a large pressure and temperature dependence.
- Density reduction by hydrogenation of hcp-Fe is larger than that of fcc-Fe.

## Abstract

Hydrogen is a promising candidate for the "light" element in terrestrial planetary cores. Its incorporation into iron causes significant volume expansion, leading to a substantial density deficit. Although extensive studies have been conducted on iron hydride ($FeH_x$) with the fcc structure, the thermoelastic properties on $FeH_x$ with hcp structure (hcp-$FeH_x$) remain unconstrained because of the experimental difficulties to control hydrogen content. Here, we synthesized hcp-$FeH_x$ with controlled hydrogen contents under high-pressure and high-temperature conditions. We carried out *in situ* X-ray diffraction measurements on hcp-$FeH_x$ at 10–25 GPa and 300–900 K using a Kawai-type multianvil apparatus and constructed their equations of state. By combining our results with previously reported equations of state for hcp-Fe and experimental determinations of hydrogen content in hcp-$FeH_x$, we demonstrated that the discrepancies in the hydrogen-induced volume expansion coefficient can be clearly explained by its pressure and temperature dependence. Our results revealed that the hydrogen-induced volume expansion of hcp-Fe exhibits a strong temperature dependence at low pressures, but its temperature effect significantly weakens with increasing pressure. We also showed that the density reduction of Fe by hydrogenation depends on its crystal structure. These findings demonstrate that estimates of hydrogen content in iron at planetary interior conditions based on hydrogen-induced volume expansion need to be revised by properly accounting for its *PT*-dependence and crystal structure.

## 1. Introduction

Hydrogen, the most abundant element in the universe, can be a primary light element in planetary cores [Hirose et al., 2022; Shahar et al,.2026] because of its siderophile feature to iron under high-pressure and high-temperature (high-*PT*) conditions [e.g., Okuchi, 1997; Tagawa et al., 2021]. Its partition coefficient between solid and liquid iron is relatively high (0.5–0.8) among the light element candidates [Alfè et al., 2002; Oka et al., 2022; Yuan and Neumann, 2023; Zhang, Z. et al., 2025], and a notable amount of hydrogen in the whole core can be supplied to the inner core during core solidification. However, the hydrogen content in the planetary interiors remains a wildcard because the solubility of hydrogen in iron is very low at atmospheric pressure [Da Silva and Mclellan, 1976]. This characteristic prohibits the access to the estimation of the hydrogen content in the meteorites and the recovered sample from high-pressure experiments. The solubility of hydrogen in iron drastically increases at ~3 GPa and can form stoichiometric iron hydride [e.g., Fukai and Suzuki, 1986; Badding et al., 1991]. It indicates the necessity to conduct *in-situ* measurements of iron hydrides at high-*PT*.

Hydrogenation of iron modifies some physical properties: phase diagram, melting point, and magnetism [Sakamaki et al., 2009; Gomi et al., 2018]. One of the characteristics of hydrogenation of iron is a significant expansion of the unit-cell volume. Among the various physical effects associated with iron hydrogenation, volume expansion is one of the most pronounced and experimentally accessible quantities. Hydrogen incorporates into the interstitial sites of the close-packed structure of Fe such as face-centered cubic (fcc) and hexagonal close-packed (hcp) structures. Hydrogenation expands the volume and, the density of iron decreases substantially. Thus, quantification of the volume expansion by iron hydrogenation is essential to evaluate the relation between the hydrogen content and the density in the planetary cores. The relation between hydrogen content and volume expansion can be described as the hydrogen-induced volume expansion per one H atom ($v_{\mathrm{H}}$) as follows.

$$v_{\mathrm{H}} = \frac{V_{\mathrm{FeH}_x} - V_{\mathrm{Fe}}}{\mathrm{Z}x}, \quad (1)$$

where $V_{\mathrm{FeH}_x}$ and $V_{\mathrm{Fe}}$ are the unit-cell volumes of $\mathrm{FeH}_x$ and Fe at a given *PT* condition. $x$ and Z represent the amount of absorbed hydrogen and the number of host atoms in the unit cell, respectively. Hydrogen content ($x$) in $\mathrm{FeH}_x$ can be estimated from the structure refinements of neutron diffraction (ND) profiles [Antonov et al., 1998; Machida et al., 2014; Machida et al., 2019; Ikuta et al., 2019]. By quantitatively characterizing the substantial volume increase caused by hydrogenation, hydrogen contents can be inferred directly from the unit-cell volume.

Measuring this physical quantity is relatively straightforward in polycrystalline samples under high-*PT* conditions and is robust against texture development and grain growth.

Within the *PT*-range accessible in laboratory experiments on the Fe–H systems containing sufficient hydrogen fluid, either dhcp- or fcc-structured iron hydrides are stabilized [Sakamaki et al., 2009; Kato et al., 2020], and even superhydrides can form at very high pressure [Pépin et al., 2017]. Most experimental studies predominantly focused on hydrogen-rich compositions (~18 wt% H). Previous investigations of hydrogen-induced volume expansion under high-*PT* conditions predominantly focus on fcc-structured $FeH_x$ (fcc-$FeH_x$). The reported values of $v_H$ of fcc-$FeH_x$ are consistently ~2.2 Å$^3$ per H atom [Antonov et al., 1998; Machida et al., 2014; Ikuta et al., 2019]. Recent X-ray diffraction (XRD) work has examined the *PT*-dependence of $v_H$ [Tagawa et al., 2022a].

Meanwhile, phase studies of Fe–H indicated that hydrogen incorporation expands the stability field of the hcp structure to higher temperatures, especially above 30 GPa [Oka et al., 2022; Tagawa et al., 2022b]. Consequently, iron hydride with hcp structure can be the important phase in planetary bodies with core pressures to those in Mercury-, and Mars-size planetary bodies. Therefore, the physical properties of hcp-$FeH_x$ represent essential parameters for understanding not only the Earth's core but also for establishing constraints on hydrogen storage in planetary cores across a wide range of terrestrial bodies.

Despite its geophysical importance, experimental studies on $FeH_x$ with the hcp structure (hcp-$FeH_x$) remain very limited because of the experimental difficulties in precisely controlling the hydrogen content in the system. Producing single-phase hcp-$FeH_x$ requires precise control of both confining and hydrogen pressures. Because of the limited experimental investigations of hcp-$FeH_x$, two major issues regarding volume change associated with hydrogen incorporation remain at current stage. First, the *P–V–T* relation of $FeH_x$ with the hcp structure have yet to be explored experimentally. Although a previous computational study reported equations of state for hcp-$FeH_x$, its thermoelastic properties were assumed to be identical to those of pure hcp-Fe [Caracas, 2015]. Second, the $v_H$ for the recovered sample at 90 K [Antonov et al., 1998] is 20–30% smaller than that obtained later at ~4–7 GPa up to 1073 K [Machida et al., 2019]. While the $v_H$ values obtained in the two previous studies fall within the range of transition metal hydrides [Fukai, 2006], such discrepancies introduce a large uncertainty in estimating hydrogen content of planetary cores from density [Mori et al., 2024].

Recent progress in the Kawai-type multi-anvil assemblies (MA6-8) system for high-pressure and high-temperature neutron diffraction [Sano-Furukawa et al., 2021] has increased the pressure limit of high-*PT* neutron diffraction [Mori et al., 2021; 2024]. However, the lower neutron flux compared to synchrotron X-ray diffraction, even using pulsed sources, limits structural refinements under high-*PT* conditions. The discrepancies in $v_H$ likely to originate from the difference in the *PT* conditions of the neutron diffraction experiments because the numerator on the right-hand-side of Eq.1 is indeed a function of at least two variables. Thus, Eq.1 can be rewritten as a function of pressure (*P*) and temperature (*T*) as follows,

$$v_{\mathrm{H}}(P,T) = \frac{V_{\mathrm{FeH}_x}(P,T) - V_{\mathrm{Fe}}(P,T)}{\mathrm{Z}x}, \quad (2)$$

where $V_{\mathrm{FeH}_x}(P,T)$, $V_{\mathrm{Fe}}(P,T)$, Z, and $x$ are the unit-cell volume of iron hydride, the unit-cell volume of iron, the number of atoms in the unit-cell, and the hydrogen content.

In this study, we determine the equation of state (EoS) of hcp-$FeH_x$ and elucidate its *PT*-dependence of $v_H$. Establishing the EoS of hcp-$FeH_x$ and its $v_H$ (*P*, *T*) therefore provides a practical framework for estimating hydrogen contents in iron, offering a critical link between high-*PT* experiments on Fe–H systems and the density deficits observed in planetary cores.

**2. Experimental Design and Procedure**

High-*PT* X-ray diffraction experiments were performed at BL04B1 of the Synchrotron Radiation Facility, SPring-8 [Utsumi et al., 1998]. Pressure was generated using the MA6-8 configuration with a DIA-type guide block and a large volume press, SPEED-Mk.II [Katsura et al., 2004]. The schematics of pressure generation and the details of the cell assembly are shown elsewhere [Takano et al., 2026]. We used an octahedral $Cr_2O_3$-doped MgO pressure medium with an edge length of 10 mm. This octahedral pressure medium was set in the center of the inner eight second-stage anvils (MA8). Tungsten carbide anvils with 26 mm on one side and 4 mm truncation edge lengths were used for the second-stage anvils. This Kawai-type cell was compressed via the outer six first-stage anvils. During heating, the function generator controls the voltage. The voltage is controlled by sending commands to the function generator to keep the power constant via software [Katsura et al., 2004]. Temperature was monitored using a W3%Re–W25%Re thermocouple every 10 seconds. No correction was made for the effect of pressure on electromotive force.

As a starting material, iron powder (FUJIFILM Wako Pure Chemical Corporation, Japan) was mixed with hexagonal BN (hBN) in a 2:3 volume ratio to inhibit grain growth. The sample was packed in a NaCl container. NaCl is empirically an excellent material for a hydrogen-sealing capsule up to the phase transition from the B1 to the B2 structure [Matsuoka et al., 2019]. Inside the NaCl container, a pelletized sample and a disk of $NH_3BH_3$ hydrogen source, which decomposes irreversibly at high temperatures [Nylén et al., 2009; Kakizawa et al., 2021], are loaded. Since $NH_3BH_3$ is a solid hydrogen source, the amount of hydrogen released can be controlled by varying the weight ratio of the hydrogen source to the sample. In this study, the amount of hydrogen released was adjusted to approximately 0.35 relative to the Fe ratio and sealed in the container. A cylindrical $TiB_2$+AlN+hBN composite heater [Kanzaki, 2010] was used to generate high temperatures. All XRD measurements were performed while the entire press was oscillated along the $\kappa$ axis between 0° and 6° with respect to the direction of the incident X-rays to eliminate the grain growth effect on the XRD profiles. Pressure was determined from the EoS of NaCl [Matsui et al., 2012] in the NaCl+MgO plate inserted directly under the thermocouple.

We first compressed the sample to ~12 GPa and heated it to 800–900 K to decompose $NH_3BH_3$ and hydrogenate hcp-Fe. To avoid the formation of fcc-$FeH_x$, hydrogenation was carried out by increasing temperature stepwise. At the pressures investigated in this study, the hcp–fcc phase boundary of pure iron lies at approximately 800–900 K [e.g., Yamazaki et al., 2012] and hydrogen incorporation into hcp-Fe increases the temperature of the hcp–fcc phase boundary. Preliminary experiments showed that once fcc-$FeH_x$ forms, it can be preserved as a metastable phase upon cooling because of kinetic effects. The coexistence of hcp and fcc phases would induce hydrogen partition between those phases, preventing quantitative evaluation of hydrogen content and reliable determination of EoS because the hydrogen solubility limit in fcc-$FeH_x$ is significantly higher than that in hcp-$FeH_x$ [Mita et al., 2025].

Time-resolved X-ray diffraction profiles (2 minutes per profile) were collected to confirm the completion of hydrogenation based on volume expansion (Fig. 1). After that, $V(P, T)$ was determined by sequential XRD measurements as the temperature was gradually decreased from 900 K to 300 K with a constant load for 2 hours. Although this approach has a shortcoming in constructing isothermal compression curves at high temperature, a detailed unit-cell volume data set can be obtained with respect to temperature without significant pressure deviation by slow cooling. Preliminary experiments with these assemblies have shown that the pressure decreases linearly with decreasing temperature once the cell has been heated. In this experiment, pressure was measured at ~600 K, halfway in cooling to room temperature (the solid gray circles in Fig. 2(a)). Each pressure measurement was performed with a 4-minute exposure time. Those *PT* conditions were along a straight line between the pressures measured at 900 K and 300 K. Thus, the assumption that the *P–T* relationship during the cooling process at constant load is linear is well established for this cell. The sample was cooled from 900 K to an intermediate temperature (~600 K) for 1 hour, then cooled to 300 K for another 1 hour.

Each sequential XRD profile was collected for 2 minutes in a descending-temperature path (~60 frames were obtained from 900 K to room temperature). The pressure and temperature change during a single XRD profile acquisition (the difference between maximum and minimum temperature and the estimated pressure during the single data collection) were $\Delta T$~15 K and $\Delta P$~0.05 GPa, respectively. Those *PT* deviations are small enough for iron to settle the ordinal EoS from the *P–V–T* data. Note that the amount of hydrogen in hcp-$FeH_x$ synthesized under hydrogen-unsaturated conditions is assumed to be invariant during the experiment. The obtained *P–V–T* dataset was fitted using Eosfit7 software [Gonzalez et al., 2016]. We also conduct a hydrogenation experiment and monitor the unit-cell volume of hcp-$FeH_x$ throughout compression and during multiple heating and cooling cycles. Detailed experimental procedure and its analysis are shown in Appendix. X-ray diffraction profiles were analyzed by fitting diffraction peaks and refining the lattice constants of both the sample and the pressure marker using the least squares method implemented in PDIndexer software [Seto et al., 2010; 2025].

## 3. Results and Discussion

### 3.1 Elastic behavior of hcp-$FeH_x$ at high-*PT* conditions

First, we obtained hcp-$FeH_x$ at 12 GPa and 900 K. During hydrogenation at 900 K, the unit-cell volume of hcp-Fe was monitored using time-resolved XRD measurements. The unit-cell volume of $FeH_x$ gradually expanded by hydrogenation and reached a steady state after ~1 hour (Fig. 1). After that, we performed sequential XRD measurements along the cooling paths at five constant loads. The *PT* conditions and unit-cell volumes are listed in [Mori, 2025], and Fig. 2(b) shows the *P–V* plots. Physical properties, such as electrical properties and slip systems, of hcp metals correlate with the axial ratio *c*/*a*. Figure S1 shows the axial ratio *c/a* of hcp-$FeH_x$ under high-*PT* conditions. It increases with temperature and decreases with pressure, consistent with pure hcp-Fe [Yamazaki et al., 2012]. Notably, the *c/a* of $FeH_x$ is slightly larger than that of pure Fe under the same conditions, indicating that hydrogenation increases the axial ratio. Although theoretical studies suggest possible ferromagnetism in hcp-$FeH_x$ [Gomi et al., 2018], no clear anomalies from spin transitions [Ono et al., 2015] or magnetic contribution were observed, implying negligible magnetic effects in our *P–T* range.

The obtained *P–V–T* dataset of hcp-$FeH_x$ was fitted to Birch-Murnaghan (BM) EoS as follows,

$$P(T_0, V) = \frac{3}{2}K_0\left[\left(\frac{V_0}{V}\right)^{\frac{7}{3}} - \left(\frac{V_0}{V}\right)^{\frac{5}{3}}\right]\left[1 + \frac{3}{4}(K_0' - 4)\left\{\left(\frac{V_0}{V}\right)^{\frac{2}{3}} - 1\right\}\right]. \quad (3)$$

The parameters, $K_0$, $K_0'$,, and $V_0$ are the isothermal bulk modulus, its pressure derivative, and the standard unit-cell volume, respectively. Because of the narrow pressure range in our dataset, it is insufficient to derive the pressure derivative of the bulk modulus.
Caracas (2015) reported the EoS on hcp-$FeH_x$ with various *x* values by the first-principles calculations: $K_0' = 4.339$ (hcp-$FeH_{0.25}$) and $K_0' = 4.227$ (hcp-$FeH_{0.5}$). In this study, we assumed that $K_0'$ is equal to 4.3. We also fitted the obtained dataset to Vinet equations of state. The resultant compression behavior yields nearly the same as that fitted to BM EoS.

To describe the high-temperature compression behavior of hcp-$FeH_x$, we fitted the dataset using the Mie–Grüneisen–Debye (MGD) thermal equation of state under the quasi-harmonic approximation (QHA). This model is preferable for extrapolation to low temperatures, such as those in Antonov et al. (1998). Within the framework of the QHA assumption, the vibrational modes of the solid are assumed to be isotropic because the ratio of the linear thermal expansion to the compressibility is required to be isotropic. Materials with the hcp structure generally exhibit anisotropic elastic properties. Similarly, in this case, the lattice constants *a* and *c* obtained along each cooling path showed anisotropic behavior (Fig. S2). Although this assumption is also not strictly valid for hcp-Fe, it is often used as a practical simplification when describing its EoS [Uchida et al., 2001; Yamazaki et al., 2012; Fei et al., 2016]. To compare those previous EoS studies, we applied the MGD model to derive the thermal effect. For the thermal EoS, pressure is described as

$$P(V, T) = P_{T_0}(V) + \Delta P_{th}(V, T), \quad (4)$$

where $P_{T_0}(V)$ and $\Delta P_{th}(V,T)$ are the pressure at standard temperature of $T_0$ (which is 300 K in this study) and the thermal pressure, respectively. Using the Mie–Grüneisen relation, thermal pressure can be written as

$$\Delta P_{th} = (E - E_0)\frac{\gamma}{V}, \tag{5}$$

where $\gamma$ is Grüneisen parameter. Applying the Debye model, the thermal-energy term can be described as

$$E = 9nRT\left(\frac{T}{\theta}\right)^3 \int_0^{\frac{\theta}{T}} \frac{z^3}{e^z - 1}\mathrm{d}z, \tag{6}$$

and

$$E_0 = 9nRT_0\left(\frac{T_0}{\theta}\right)^3 \int_0^{\frac{\theta}{T_0}} \frac{z^3}{e^z - 1}\mathrm{d}z, \tag{7}$$

where $R$, $n$, and $\theta$ are the gas constant, the number of atoms per formula unit, and the Debye temperature, respectively. Since $n$ in Eqs. 6 and 7 is the number of atoms per formula unit, $n$ is equal to $1+x$ in this case. However, in metal hydrides, the mass ratio of metal to hydrogen is very huge, and the dissolved hydrogen can be ignored. The following empirical assumptions are introduced to determine the volume dependence of the Grüneisen parameters:

$$\gamma = \gamma_0\left(\frac{V}{V_0}\right)^q, \tag{8}$$

where $\gamma_0$ and $q$ are the standard Grüneisen parameter and dimensionless parameter, respectively. Debye temperature ($\theta$) can be described as

$$\theta = \theta_0 \exp[(\gamma_0 - \gamma)/q], \tag{9}$$

using the standard Debye temperature, $\theta_0$. Volume dependence of $\gamma$ and $\theta$ was modelled empirically. The fitted parameters are listed in Table 1(a). Anharmonic and electronic contributions were neglected, as their total effects are estimated to be ~1% of the QHA thermal pressure (Fig. S3). The Debye temperature of hcp-$FeH_x$ was ~500–700 K. This value is lower than that of pure hcp-Fe obtained by *PVT* experiments with a Kawai-type multianvil apparatus (KMA), but larger than the commonly used value, 420 K [Anderson et al., 2001; Shen et al., 2004]. Another approach for describing the thermoelastic effect is to apply the empirical expression for the thermal expansion coefficient. In this case, the results differ slightly from the MGD model, but the trend is not significantly different (See Text S1 and Table S1).

In contrast, EoS for hcp-Fe shows different trends depending on the study. The choice of EoS for hcp-Fe significantly affects the *PT* dependence of $v_H$ because $V_{Fe}$ at a certain condition is calculated using the EoS for hcp-Fe (Table 1). Reports on the compression behavior of hcp-Fe at high temperature is limited. We applied five thermal EoS for hcp-Fe: Four are based on *PVT* studies [Uchida et al., 2001; Yamazaki et al., 2012; Fei et al., 2016; Miozzi et al., 2020] and one is a recent computational result [Zhang, Y. et al., 2025]. Table 1(a) lists the elastic parameters of hcp-$FeH_x$ in comparisons with that of hcp-Fe. The established EoS of hcp-$FeH_x$ is in good agreement with the measured unit-cell volume, with a difference of 0.2% in volume (Fig. S4). The thermal pressure of hcp-$FeH_x$ at lower pressures is significantly large compared to that of pure Fe with hcp structure but drastically decreases with shrinking the unit-cell volume (compression) and approaches that for pure iron (Fig. S5).

### 3.2 Hydrogen content ($x$) in hcp-$FeH_x$ through the measurements

By comparing EoS for hcp-$FeH_x$ with that of hcp-Fe, the 'relative' value of $v_H(P, T)$ can be derived. To determine the 'absolute' value of $v_H(P, T)$, hydrogen content ($x$) estimation in hcp-$FeH_x$ is required. The scattering cross section by X-ray depends on the number of electrons. Thus, a change in hydrogen concentration in hcp-$FeH_x$ is almost undetectable from the obtained XRD profile. To date, hydrogen content ($x$) and the unit-cell volume of hcp-$FeH_x$ ($V_{\mathrm{FeH}_x}$) have been determined simultaneously at three distinct *PT* conditions by neutron diffraction experiments. Letting the pressure and temperature conditions, where the neutron diffraction profile for the crystal structure refinements was obtained, be $(P_i, T_i)$. Combined with EoS for hcp-Fe, $v_{\mathrm{H}}(P_i, T_i)$ can be determined by Eq. 2. By substituting $(P_i, T_i, v_{\mathrm{H}}(P_i, T_i))$ into Eq. 2, hydrogen content ($x$) of hcp-$FeH_x$ can be estimated.

*PT*-dependence of $v_H$ varied with the choice of EoS for hcp-Fe (Figs. 3, S6, and S7). Changing the hcp-Fe EoS alters the estimation of hydrogen content by up to ~15%, and $x$ yields values of ~0.30–0.34. Averaging 15 values (5 EoS×3 $(P_i, T_i)$ conditions), the hydrogen content of hcp-$FeH_x$ examined in this study for constructing EoS is $x$ = 0.32(1) (parentheses indicating the standard deviation of hydrogen content at each condition). Notably, the hydrogen content across three $(P_i, T_i)$ conditions remains nearly constant within a given EoS, with variations of less than 5%. This internal consistency strongly supports the conclusion that the hydrogen content in hcp-$FeH_x$ can be fixed regardless the choice of EoS (Table 1(b)), and that the *PT*-dependence in $v_H$ is the origin of the discrepancies of $v_H$ derived from neutron diffraction experiments. We also investigated variations in hydrogen content during compression and with temperature change using our derived $v_H(P, T)$. The results showed consistent unit-cell volume during heating and cooling, and no significant scatter in hydrogen content during compression (see Appendix). This observation supports the assumption that the hydrogen content remained nearly constant within our experimental resolution. Conversely, our results demonstrated that $v_H$ of hcp-$FeH_x$ is a function of both pressure and temperature, and the discrepancies in $v_H$ values can be reconciled by explicitly considering its *PT* dependence.

### 3.3 $v_H$ ($P$, $T$) of hcp-$FeH_x$

The derived $v_H$ ($P$, $T$) of hcp-$FeH_x$ are shown in Fig. 3 (Fig. S6 was constructed using the EoS of hcp-$FeH_x$ with the dimensionless parameter $q$ set to 1). The ydrogen content was fixed to 0.32 and five thermal EoS of hcp-Fe were applied. The $P$–$v_H$ curves show the general feature of pressure-induced shrinkage of $v_H$. This pressure dependent behavior is consistent with earlier *ab initio* calculations by Skorodumova et al. (2004) and also with experiments for the other transition metals adjacent to Fe, such as Co and Ni [Fukai, 2006]. The temperature effect on $v_H$ is positive, but its effect diminishes with increasing pressure. The *PT*-dependent $v_H$ behavior is robust regardless of the models. Figure 4(a) shows the extracted $P$–$v_H$ relation at 300 K and 1000 K. The uncertainty arises from the difference in the EoS of hcp-Fe, capturing the aforementioned *PT*-effect on $v_H$ beyond the model uncertainty. In this context, it is problematic to directly use $v_H$ values estimated from neutron diffraction experiments to calculate possible hydrogen contents in planetary cores based on density deficits. For example, the value of $v_H$ at 5 GPa and 600 K, where high-*PT* neutron diffraction measurements were conducted by Machida et al. (2019), is approximately 50% larger than that at 100 GPa at the same temperature. Direct extrapolation of $v_H$ values obtained at low-pressure and high-temperature conditions therefore leads to underestimation of the hydrogen content at much higher pressures. Our result also indicates that pressure dominantly control the density reduction by hydrogen incorporation, whereas the temperature effect is strongly suppressed as pressure increases.

### 3.4 The difference in $v_H$ between hcp and fcc phase

The stability field of single-phase hcp-$FeH_x$ expands markedly to higher temperatures above ~30 GPa [Tagawa et al., 2022b]. To compare $v_H(P, T)$ of $FeH_x$ with hcp and fcc structures, the $v_H(P, T)$ of fcc-$FeH_x$ was estimated using EoS of nonmagnetic fcc-$FeH_x$ with $x \sim 1$ by Tagawa et al. (2022a) and EoS of pure fcc-Fe by Tsujino et al. (2014). Figure 4(b) shows the ratio between the $v_H(P, T)$ of fcc-$FeH_x$ and the $v_H(P, T)$ of hcp-$FeH_x$. In comparison with hcp-$FeH_x$, the temperature effect on $v_H$ of fcc-$FeH_x$ is small [Tagawa et al., 2022a; Zhang et al., 2024].

These results are consistent with the neutron diffraction data for fcc-$FeH_x$ [Machida et al., 2014; Ikuta et al., 2019], which shows no significant $PT$-effects within the experimental uncertainty. The observed temperature effect on $v_H$ is significantly large in hcp-$FeH_x$ than in fcc-$FeH_x$ at low pressures. Over a wide pressure range (up to ~100 GPa), the $v_H$ of hcp-$FeH_x$ is larger than that of fcc-$FeH_x$ by approximately 20–30%.

It should be noted that both experimental and computational studies suggest that the molar volume of hcp-Fe is slightly smaller than that of fcc-Fe at atmospheric pressure calculated by their EoS [e.g., Yamazaki et al., 2012; Tsujino et al., 2014]. However, this difference is small and does not result in a significant density contrast even near the fcc–hcp phase boundary, remaining below ~5% [Dorogokupets et al., 2016]. To assess the crystal-structure dependence of volume expansion, we considered a hypothetical compression behavior of hcp-$FeH_x$ with hydrogen content fixed at $x = 1$ and compared it with the previously reported compression behavior of fcc-$FeH_x$ ($x \sim 1$). The unit-cell volumes of $FeH_x$ normalized with respect to Fe shows that the volume expansion of hcp is larger than that of fcc. Hence, the molar volume of hcp-Fe is smaller than fcc-Fe, and hydrogenation results in a density reversal at a certain H concentration.

### 3.5 Geophysical implication

The crystal structure of the iron alloy in the planetary core can affect the density reduction induced by hydrogenation. Recent molecular dynamics simulations of the Fe–H binary system have estimated the maximum hydrogen concentration in the cores of rocky bodies of varying sizes [Stoutenburg et al., 2026]. Their results indicate that the maximum solubility of hydrogen in planetary cores with Mercury-like and Mars-like structures is approximately 1.5 wt% and 4.0 wt%, respectively. These concentrations exceed the eutectic composition of the Fe–H system, indicating that hydrogen content of planetary cores is highly sensitive to their thermal and chemical evolution.

Geodetic constraints suggest that Mercury has a large solid inner core [Genova et al., 2019]. The presence of a magnetic field originating from its core [Plattner and Johnson, 2011] and libration signatures [Margot et al., 2007] indicate that Mercury's core is likely partially molten. Because of those observational features, light elements are dissolving into the core, which reduces the melting temperature of iron. So far, S, Si, and C have been intensely investigated based on Mercury's surface characteristics [Shahar et al., 2026], albeit their abundance and the possibility of hydrogen incorporation into the core remain uncertain. Under the conditions at the center of Mercury and Mars (~40 GPa), Fe–X (X is a major light element candidate) binary systems of H and S (Fe–FeH and Fe–$Fe_3S$) exhibit significantly lower eutectic temperatures than those of Si and C (Fe–FeSi and Fe–$Fe_3C$) [Oka et al., 2022; Mita et al., 2025; Chudinovskikh and Boeheler, 2007; Stewart et al., 2007; Sakai and Hirose, 2026; Fischer et al., 2013; Mashino et al., 2019]. In addition, hydrogen can dissolve into fcc or hcp iron in large atomic fractions without altering its crystal structure, unlike the other Fe–X systems. This suggests that hydrogen can be incorporated into Fe-rich alloys with compositions close to pure iron, even in the presence of other light elements.

In the cooling path of the Fe–H phase diagram at low hydrogen concentrations, the crystallization of Fe–H alloy is predicted to pass through a peritectic point (*Tp*), which is higher than the eutectic point [Tagawa et al., 2022b]. Below *Tp*, hcp-$FeH_x$ stabilizes as a single stable solid phase. This behavior indicates that hydrogen expands the stability field of the hcp phase to higher temperatures. It has also been reported that silicon slightly enhances the thermal stability of the hcp phase [Fischer et al., 2017]. Although fcc-$FeH_x$ crystallizes during the earliest stage of core solidification at ~40 GPa, hcp-$FeH_x$ will appear during subsequent cooling, and its transition temperature is likely higher than that of pure iron by hydrogenation [Mita et al., 2025].

Figure 5(a) shows the relationship between hydrogen concentration and density in iron hydrides at 40 GPa, neglecting phase stability. For pure iron, the hcp phase is denser than the fcc phase. However, as discussed in the previous section, $v_H$ of hcp-$FeH_x$ is larger than that of fcc-$FeH_x$. At low hydrogen contents, the hcp phase is denser than the fcc phase, whereas the latter becomes denser by increasing the hydrogen content in $FeH_x$. As a result, a density crossover occurs at certain hydrogen concentrations. An example illustrating the effect of realistic hydrogen concentrations in the Mercury and Mercury-like core is portrayed in Fig. 5(b). Here, the *Tp* of the Fe–H system is assumed to be 2000 K based on the Fe and Fe–H phase diagrams. Crystallization of hcp-$FeH_x$ decreases the density more efficiently than fcc-$FeH_x$. In the phase loop, the system separates into H-poor fcc-$FeH_x$ and H-rich hcp-$FeH_x$. At 2000 K, this phase separation produces a density difference of approximately 2% (~0.17 g/cm$^3$) between the dense H-poor fcc-$FeH_x$ and the light H-rich hcp-$FeH_x$.

The decrease in density caused by hydrogenation can affect the estimation of the required hydrogen content to reproduce the core density. Here, we evaluate the hydrogen concentration necessary to reproduce the density at the peritectic point (0.3 wt%, $Tp$ = 2000 K). Below this temperature (and also below this H-concentration), the single phase of hcp-$FeH_x$ at the peritectic point becomes thermodynamically stable. Using our $PT$-dependent $v_H$ for hcp-$FeH_x$, the density is ~8.86 g/$cm^3$. Directly applying the density reduction per H-atom inferred from fcc-$FeH_x$ to hcp-$FeH_x$, results in the required hydrogen content is 0.5 wt%, an overestimation of approximately 60%. Even if estimating the hydrogen content using the density reduction induced by hydrogenation of fcc-Fe instead of using hcp-Fe, the hydrogen content would be 0.4 wt%, which is 22% larger. Direct application of the density reduction based on fcc-$FeH_x$ might systematically overestimate the hydrogen abundance. The coexistence of other light elements may further reduce the maximum hydrogen content required to satisfy core density constraints [Mori et al., 2024].

On the other hand, hcp-$FeH_x$ is unlikely to be a stable phase in the Martian core. The mean core density of Mars is estimated to be 5.7–6.8 g/$cm^3$ [Stähler et al., 2021; Samuel et al., 2023; Khan et al., 2023]. The possible density of the solid core at the inner core boundary is 7.2–7.7 g/$cm^3$ [Bi et al., 2025], which is lower than that of molten pure Fe under corresponding pressure conditions [Kuwayama et al., 2020]. Such a large density deficit cannot be explained by $FeH_x$ with intermediate hydrogen compositions. If iron hydrides are present in a crystallizing Martian core, the hydrogen content must exceed H/Fe > 1, and stoichiometric fcc-FeH would be the dominant phase, with no contribution from the hcp phase. A hydrogen-rich Mars core model has been proposed by Yokoo et al. (2022).

Although applying these results directly to Earth requires substantial extrapolation beyond the experimentally constrained $PT$-range, which causes substantial uncertainty, the results indicate that the difference in $v_H$ between hcp and fcc crystal structures becomes comparable at Earth's inner core. Furthermore, the difference in volume per Fe atom between those phases also converges. Therefore, the effect of the crystal structure and temperature effect on the hydrogen-induced density reduction might be negligible under the conditions of at the Earth's inner core.

## 4. Conclusion

We successfully measured the thermoelastic properties of hcp-$FeH_x$ under hydrogen-undersaturated conditions at high-*PT* conditions by taking advantage of a large-volume multi-anvil apparatus combined with a solid hydrogen source. This experimental strategy enabled precise control the hydrogen content and allowed robust *PVT* measurements of hcp-$FeH_x$, which were difficult to achieve in previous studies. We demonstrated that the previously reported inconsistent values of $v_H$ of hcp-$FeH_x$ originate from its *PT* dependence. The temperature effect on $v_H$ is pronounced at low pressures but is progressively suppressed with increasing pressure. In addition, the derived $v_H$ of hcp-$FeH_x$ is systematically larger than that of fcc-$FeH_x$ considering the *PT*-effects on both phases. These results clearly indicate that both pressure and the crystal structure of iron hydride exert a strong influence on estimates of hydrogen content in planetary cores especially with Mars- and Mercury-size bodies. It should be noted that extending the *PVT* measurements of hcp-$FeH_x$ to a broader *PT* region remains experimentally challenging. One major difficulty is controlling the hydrogen content above ~30 GPa, where hydrogenation of NaCl with the B2 structure becomes significant [Matsuoka et al., 2019]. Another limitation arises from the phase stability of iron hydrides. Within the experimental pressure range, hcp-$FeH_x$ undergoes a phase transition from hcp to fcc upon heating. It prohibits a systematic investigation of thermal effects on $v_H$. Although precise control of hydrogen content remains difficult, further studies of hcp-$FeH_x$ are strongly anticipated to generalize the hydrogen-induced density reduction at higher *PT* conditions.

**Figures and Tables**

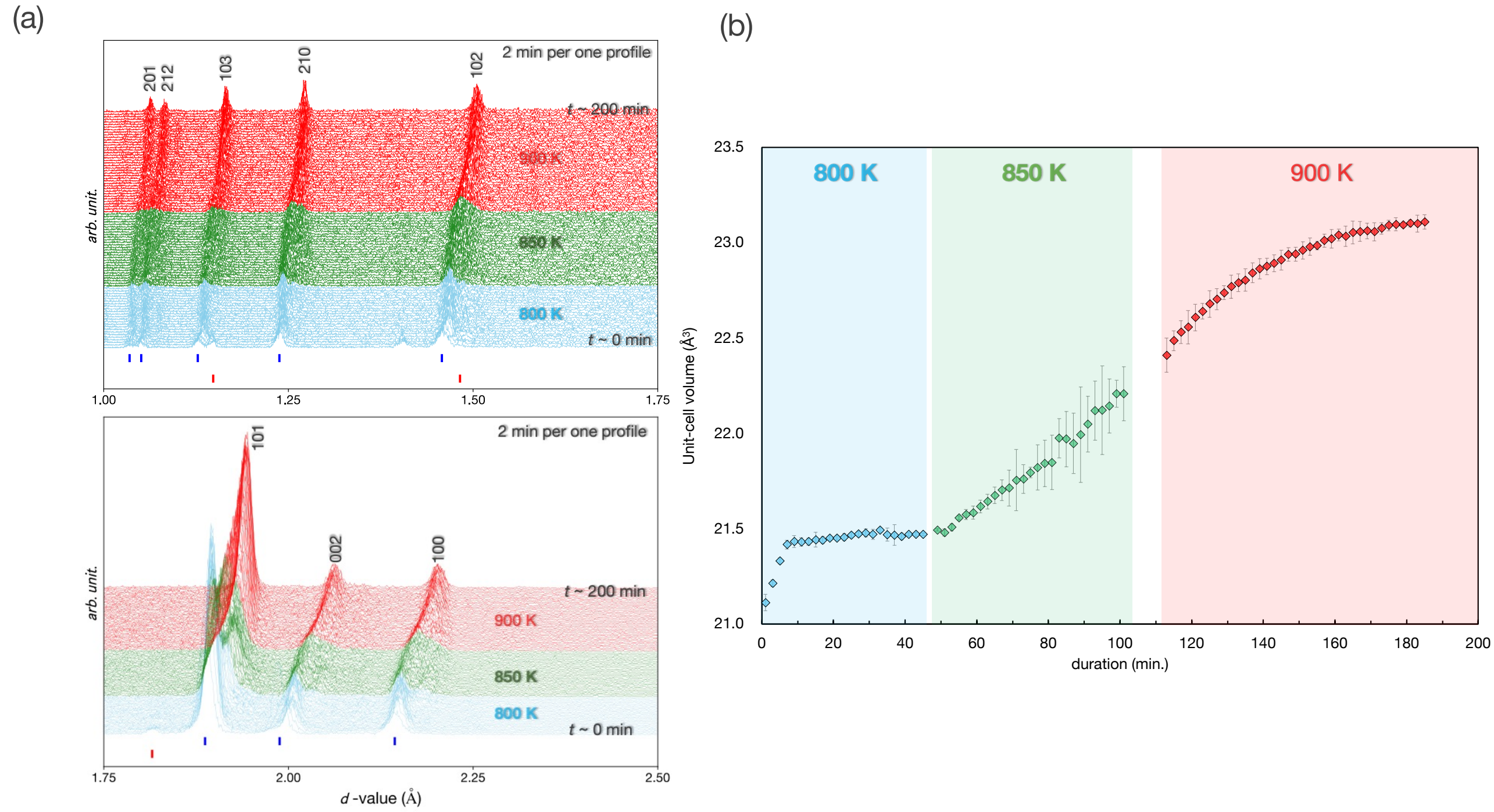

**Figure 1**: (a) Time-resolved XRD profile of hcp-FeH$_x$ during hydrogenation (from the bottom to the top). Data were obtained every 2 minutes. The color of the XRD profile indicates the temperature at which each diffraction was obtained: sky blue, green, and red corresponding to 800, 850, and 900 K, respectively. Blue and red ticks indicate the peaks of hcp-FeH$_x$ and B1-NaCl, respectively. An unknown peak emerged at 1.40 Å in the first XRD profile only. In the intermediate stage of the hydrogenation, the most intense 101 peak of hcp-FeH$_x$ was broadened and split by the contrast of the hydrogen concentration in hcp-FeH$_x$. It reflects the XRD fitting error shown in Fig. 1(b). At the final stage of hydrogenation, these peaks began to merge and eventually turned into a single sharp peak.

(b) Change in the unit-cell volume of hcp-FeH$_x$ during hydrogenation. We first heated the sample to 800 K and then increased the temperature to 900 K in steps. After approximately 60 minutes of hydrogenation at 900 K, the volume expansion almost ceased, and the volume error decreased, suggesting that hydrogenation had completed. Strictly speaking, a slight tendency for the volume to increase in the final stage of hydrogenation was observed. A complementary experiment showed that it takes ~5 hours to completely stop volume expansion, but this does not significantly affect the determination of hydrogen content ($\Delta x$~0.02 at most). See Appendix for details. We regarded that the hydrogen content in this system had reached near-equilibrium.

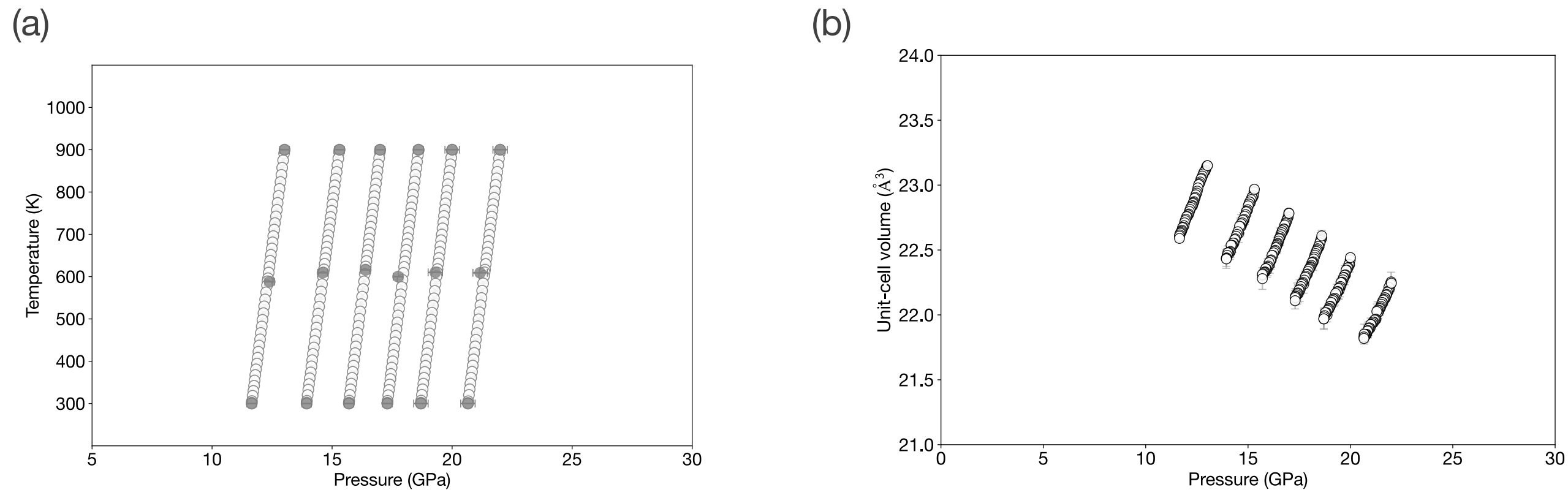

**Figure 2**: (a) *PT* conditions of collecting sequential XRD profile and *P–V* data along isotherms at 300–900 K. The grey solid circles indicate the *PT* conditions at which we collected the XRD profile of NaCl and directly determined the pressure. Those *PT* conditions fall along a straight line between the pressures measured at 900 K and 300 K (Dataset S2 in Mori, 2025). (b) The obtained unit-cell volume of hcp-$FeH_x$ at each *PT* condition.

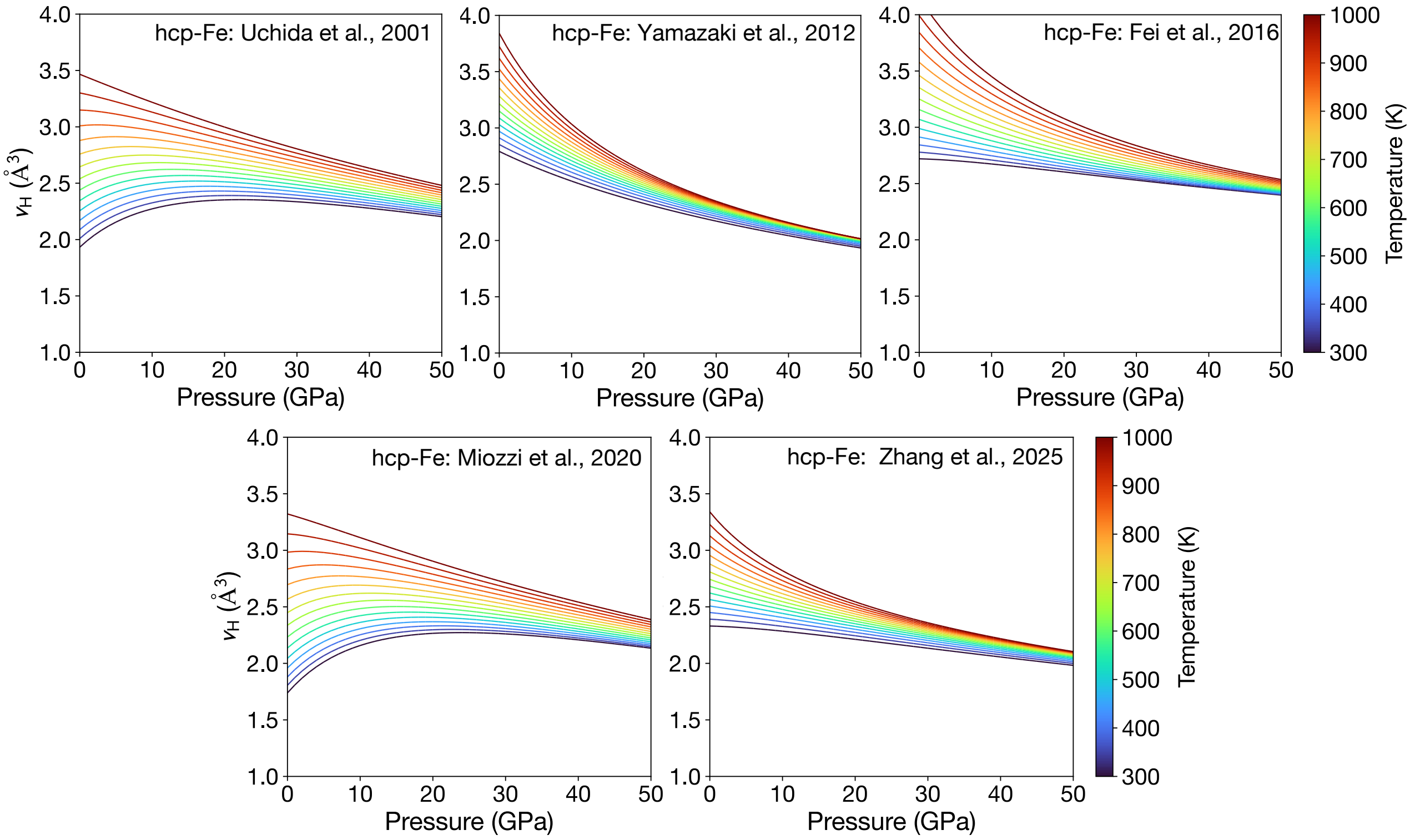

**Figure 3**: *PT* dependence of $v_H$ for hcp-$FeH_x$ at $P$ = 0–50 GPa and $T$ = 300–1000 K with various EoS for hcp-Fe. To calculate the isothermal compression behavior of hcp-$FeH_x$, we applied the BM-EoS with MGD approximation. Here, the dimensionless constant, $q$, is also optimized during fitting (*PT* dependence of $v_H$ with fixing $q$ value of EoS for hcp-$FeH_x$ to 1, which value is a commonly used value for metals, is shown in Fig. S6). Whether the value of $q$ was fixed or not has almost no effect on the chi-squared test of EoS fit.: Uchida et al. (2001), Yamazaki et al. (2012), Fei et al. (2016), Miozzi et al. (2020), and Zhang Y. et al. (2025). According to Eq. 2, $v_H$ can be derived using the unit-cell volume of hcp-Fe, $V_{Fe}$. Here, we assumed $x$ = 0.32, which is the average hydrogen content calculated across five hcp-Fe EoS.

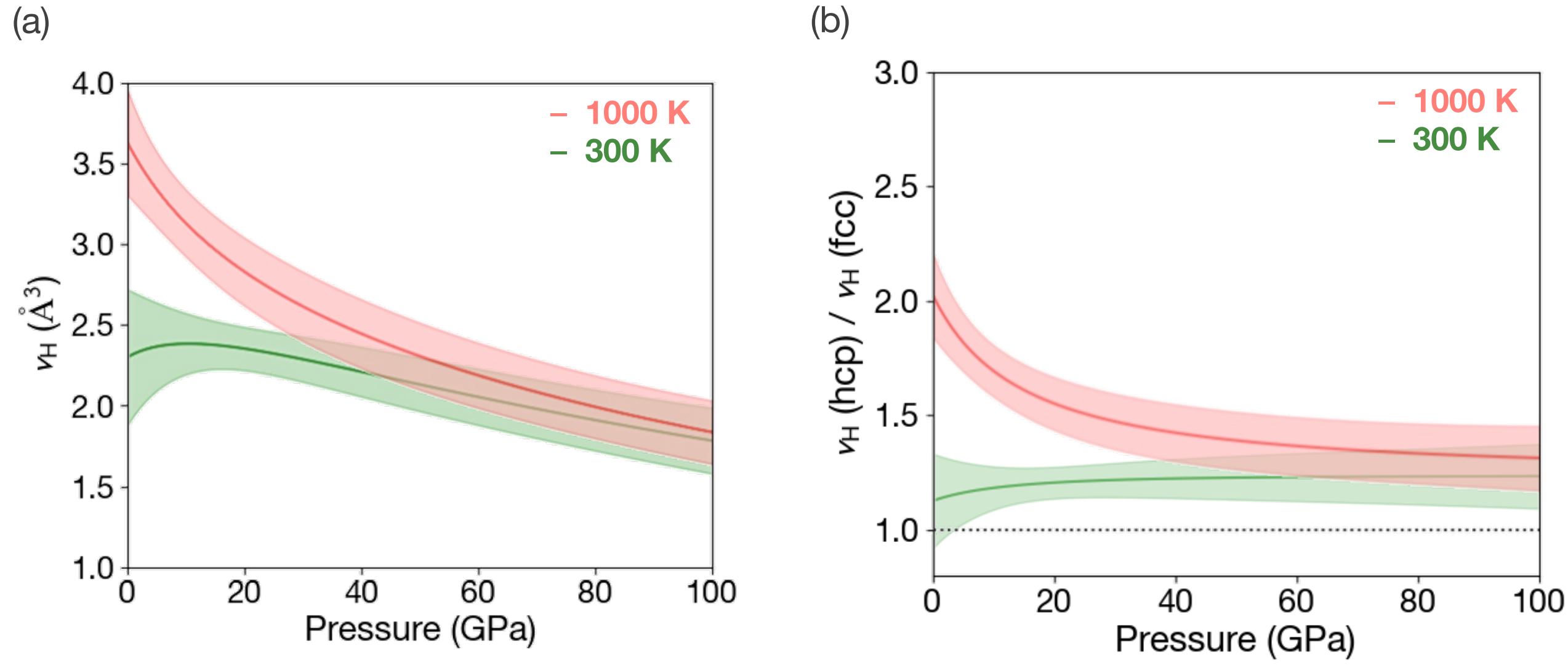

**Figure 4**: (a) The calculated isothermal curves of $v_H$ at 300 K (green solid line) and 1000 K (red solid line) up to 100 GPa. Shaded areas correspond to the standard deviation of five models at 300 K and 1000 K. (b) Ratio of $v_H$ of hcp-FeH$_x$ to that of $v_H$ of fcc-FeH$_x$ at 300 K and 1000 K as a function of pressure. The $v_H$ of fcc-FeH$_x$ were estimated using the equations of state of pure fcc-Fe [Tsujino et al., 2014] and non-magnetic fcc-FeH$_x$ with x~1 [Tagawa et al., 2022a]. The dotted horizontal line indicates unity, corresponding to the relation: $v_H$ of hcp-FeH$_x$ is equal to that of fcc-FeH$_x$

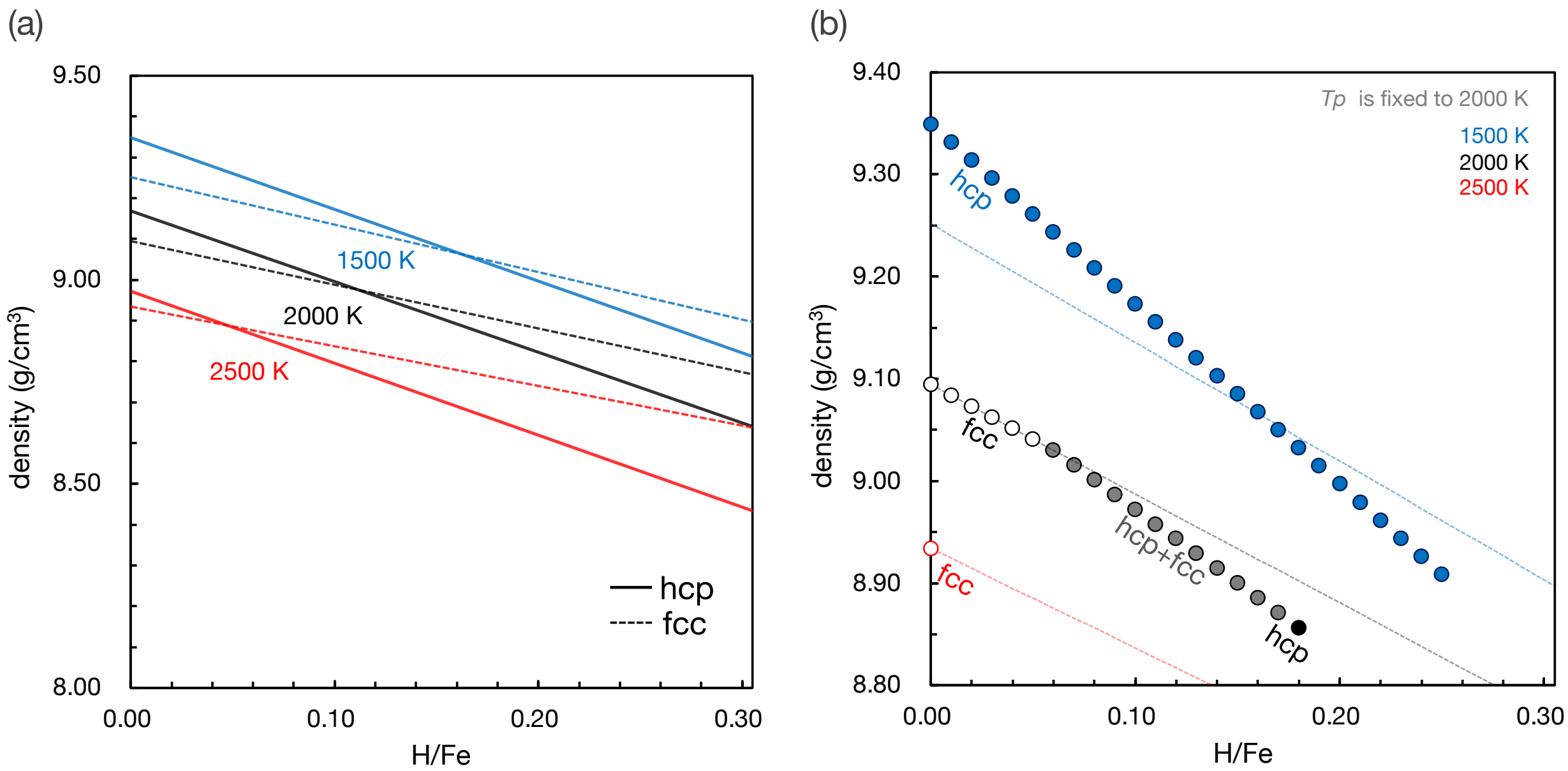

**Figure 5**: (a) Density reduction of hcp-FeH$_x$ (solid lines) and fcc-FeH$_x$ (dashed lines) as a function of hydrogen content ($x$) at 40 GPa. Line colors indicate temperature: 1500 K (blue), 2000 K (black), and 2500 K (red). Equations of state for pure hcp-Fe and fcc-Fe are taken from Yamazaki et al. (2012) and Tsujino et al. (2013), respectively. Phase stability is not considered in this figure. (b) Density reduction at 40 GPa by hydrogenation of iron based on the Fe–H phase diagram. The phase boundaries of pure Fe are assumed as follows: the melt/fcc boundary at ~2500 K [Sinmyo et al., 2019], the peritectic temperature (*Tp*) at ~2000 K [Tagawa et al., 2022b], and the hcp/fcc boundary at ~1500 K [Saxena and Dubrovinsky, 2000; Yamazaki et al., 2012]. The value of *Tp* was roughly estimated from the melting temperature of pure Fe and the eutectic temperature of the Fe–H system (~1800 K at 40 GPa) [Oka et al., 2022; Mita et al., 2025] because the eutectic temperature of the Fe–H system is not precisely determined. The circles denote the density reflecting the stable phase of FeH$_x$ (Open circles: fcc single phase; Solid circles: hcp or two-phase coexistence of H-rich hcp and H-poor fcc). Dashed lines represent the density of fcc-FeH$_x$ as a reference. At 2500 K, increasing hydrogen content leads directly to the liquidus at ~2500 K. On the other hand, at ~2000 K, which is assumed to be *Tp*, fcc-FeH$_x$ is stable at low hydrogen contents, but it separates fcc-FeH$_x$ into H-poor fcc and H-rich hcp phases above a threshold composition (the solubility limit of H in H-poor fcc-FeH$_x$). The density reduction in the co-existing region is calculated as a linear combination of the densities of those two phases, weighted by their phase fractions with the lever rule (grey solid circles). As hydrogen concentration increases, the fraction of hcp-FeH$_x$ increases, and the system reaches the solidus of hcp+Liquid shortly after becoming single-phase hcp (black solid circles). At ~1500 K, hcp-FeH$_x$ remains stable as a single phase over x = 0 – 0.3 and intersects the solidus of hcp+Liquid at higher hydrogen contents.

**Table 1**: (a) Elastic parameters of the EoS for hcp-FeH$_x$ (this study) and hcp-Fe [Uchida et al., 2001; Yamazaki et al., 2012; Fei et al., 2016; Miozzi et al., 2020; Zhang Y. et al., 2025].

(b) Calculated $v_H$, $x$ at the conditions where the structure refinements for hcp-FeH$_x$ (deuteride) using neutron diffraction were carried out [Antonov et al., 1998; Machida et al., 2019]. The two digits in the brackets of $v_H$ represent the pressure and temperature, respectively, of those experiments (e.g., $v_H$(0, 90) denotes the calculated $v_H$ at 0 GPa and 90 K). Machida et al. (2019) conducted crystal structure refinement of deuterated hcp-FeH$_x$ at 4.8 GPa and 573 K, and at 5.1 GPa and 673 K. In their paper, $v_H$ of hcp-FeH$_x$ at 5 GPa and ~500–600 K is 2.48(5) Å$^3$, which is somewhat different from our calculated value. It seems the inconsistency came from the differences in the EoS model. To reproduce their value, the EoS for pure hcp-Fe must be derived using the polynomial thermal expansion model. On the other hand, we calculated $v_H$ by applying BM-EoS with MGD approximation. The details of the polynomial thermal expansion model are described in Text S1, and we briefly present the results for this case in the Supporting Information.

(a)

| | $V_0$ (Å$^3$) | $K$ (GPa) | $K'$ | EoS | $\gamma_0$ | $\theta_0$ | $q$ | |
|---|---|---|---|---|---|---|---|---|
| FeH$_x$ | 23.94(2) | 185(3) | 4.3(fixed) | BM | 3.71(16) | 475(163) | 3.1(8) | This study |
| FeH$_x$ | 23.98(1) | 180(2) | 4.3(fixed) | BM | 3.48(13) | 669(103) | 1.0(fixed) | This study |
| FeH$_x$ | 23.93(2) | 185(3) | 4.3(fixed) | Vinet | 3.72(16) | 470(164) | 3.2(8) | This study |
| Fe | 22.7(3) | 135(19) | 6.0(4) | BM | 1.36(8) | 998(85) | 0.91(7) | Uchida et al. (2001) |
| Fe | 22.15(5) | 202(7) | 4.5(2) | BM | 3.2(2) | 1173(62) | 0.8(3) | Yamazaki et al. (2012) |
| Fe | 22.15(5) | 172.7(14) | 4.79(5) | Vinet | 1.74 | 422 | 0.78 | Fei et al. (2016) |
| Fe | 22.82(7) | 129(1) | 6.24(4) | BM | 1.11(1) | 420 | 0.3(3) | Miozzi et al. (2020) |
| Fe | 22.45(3) | 174.7(17) | 4.790(14) | BM | 2.86(10) | 1209(73) | 0.84(5) | Zhang Y. et al. (2025) |

(b)

| EoS (Fe) | $v_H^{ND}(0, 90)$ | $v_H^{ND}(4.8, 573)$ | $v_H^{ND}(5.1, 673)$ | $x^{XRD}$ (This study) |
|---|---|---|---|---|
| Uchida et al. (2001) | 1.75 | 2.60 | 2.73 | 0.315(6) |
| Yamazaki et al. (2012) | 2.42 | 2.85 | 2.93 | 0.326(6) |
| Fei et al. (2016) | 2.43 | 2.92 | 3.03 | 0.331(4) |
| Miozzi et al. (2020) | 1.66 | 2.49 | 2.63 | 0.309(5) |
| Zhang Y. et al. (2025) | 2.03 | 2.67 | 2.71 | 0.314(5) |

**Appendix: Quantification of hydrogen content in hcp-FeH$_x$**

We carried out the supplementary experiments (M4614) using X-ray diffraction at high PT but under other conditions (hydrogen content and *PT* paths). The experimental procedure and the resultant implications for the main manuscript are as follows.

The sample was compressed to ~13.3 GPa, just above the phase boundary between bcc and hcp-Fe. The obtained XRD showed that Fe partially transforms from bcc to hcp (compression did not completely transform bcc to hcp just above the phase boundary by kinetic effects). With increasing temperature, the remaining bcc phase fully transforms into the hcp phase at ~600 K. The phase boundary of hcp and fcc for pure Fe at ~12–14 GPa is ~800–900 K [Saxena and Dubrovinsky, 2000; Yamazaki et al., 2012], and hydrogenation of Fe shifts this phase boundary to a higher temperature. On the other hand, the empirically determined decomposition temperature of $NH_3BH_3$ is ~800–900 K (At this temperature, approximately two-thirds of the total hydrogen content of $NH_3BH_3$ is released as hydrogen fluid). In the additional experiment, we first heated the sample to 800 K. However, even after 50 minutes, almost no volume expansion was observed, and the unit-cell volume is the same as that of pure Fe. The pre-heating pressure in this experiment is slightly higher than that in the *PVT* measurements (M4339). This observation suggests that the decomposition temperature had not been reached or that hydrogen release required significant time even after reaching it. The temperature was increased to 900 K during the hydrogenation process and pronounced volume expansion was observed. The synthesized FeH$_x$ shows an hcp structure. As observed in the first experiment, the unit-cell volume increases very slightly at the end of the significant hydrogenation process (see Fig. 1(b) in the manuscript). In this additional run, we kept the temperature until the unit-cell volume of hcp-FeH$_x$ ceased the expansion, and it took 5 hours (Fig. A1). In the final one hour, the observed unit-cell volume is almost constant during hydrogenation.

One possible explanation for this slow volume increase is a gradual decrease in pressure during heating. Heating and decomposition of the hydrogen source cause deformation of the cell and can change pressure, as indicated by the stroke change. However, this subtle but long-lasting volume expansion kept on going much longer (several hours) compared to such relaxation processes. Although this timescale resembles the order of super-abundant vacancy (SAV) formation, it does not what we observed, because this process leads to volume contraction under high hydrogen concentration conditions. Another possibility is that a very small amount of hydrogen is slowly supplied from the hydrogen source. However, this phenomenon is predicted to be trivial. For example, to generate the difference $\Delta x \sim 0.1$, the volume change of 0.6 Å$^3$ will be required. Hydrogenation behavior observed in the EoS experiments (M4339) shows an earlier and more gradual cessation of expansion compared to this supplementary experiment (M4614) (Figs. 1(b) and A1). It indicates that volume expansion in M4339 would cease sooner than in M4621, suggesting that the sample was closer to equilibrium over the same duration. Even if the EoS experiments had exhibited the behavior similar to the supplemental one, the possible increase after apparent cessation of hydrogenation in Fig. 1(b) would be at most ~0.2 Å$^3$, corresponding to a hydrogen content change in 0.035. After confirming the completion of volume expansion, the calculated hydrogen content using our $v_{\mathrm{H}}(P, T)$ is $x$~0.23. Because the volume per metal atom increases almost linearly with H concentration [Fukai, 2006] and $v_{\mathrm{H}}$ is independent of hydrogen content ($x$), the obtained $v_{\mathrm{H}}(P, T)$ can be applied to hcp-$FeH_x$ with the other content of hydrogen.

We also measured the unit-cell volume sequentially along heating and cooling paths. Unfortunately, the thermocouple failed during hydrogenation, so the power-temperature calibration was used to estimate the temperature rather than using emf values. Generally, the temperature increase is expressed as a polynomial function of the supply input power. To create Fig. A2, we used the power-temperature relationship from the first pressure path as a provisional estimate. We should note that the actual temperature of the cell is ~750–800 K, which may be due to others using this heater ($TiB_2$+hBN) showing that the temperature generated at a given input power tends to decrease as the load (pressure) increases. On the other hand, given that the temperature at a constant load is a function of the input power, plotting the relationship between temperature (power) and volume along the heating and cooling paths at different loads is essential for exploring possible changes in hydrogen amount with temperature. Figure A3 shows the temperature and volume ($T$–$V$) relation at three pressure ranges. At both the lowest and the highest pressures, $T$–$V$ is clearly consistent with heating and cooling paths. In the mid-pressure range (16.0–18.4 GPa), the $T$–$V$ relation of the heating path is slightly deviated from that of the cooling path. This is due to the pressure change across the heating (cooling) path. As shown in Fig. A3, the pressure at 300 K before the first heating is 0.8 GPa lower than that after cooling. Also, at a high temperature, estimated to be 900 K from the first heating power relation, the calculated pressure after the first heating is slightly lower than that after the second heating. If this change could originate from a change in hydrogen content, the possible change in volume is ~0.2 Å$^3$, making the hydrogen content variation more minor than $\Delta x = 0.05$. Furthermore, we obtained X-ray diffraction data from the sample and the pressure marker at multiple points during compression (Fig. A4). The calculated hydrogen content is ~0.23 at all pressure and temperature points.

This experiment validates the assumption that the hydrogen content is nearly constant during *PVT* measurements.

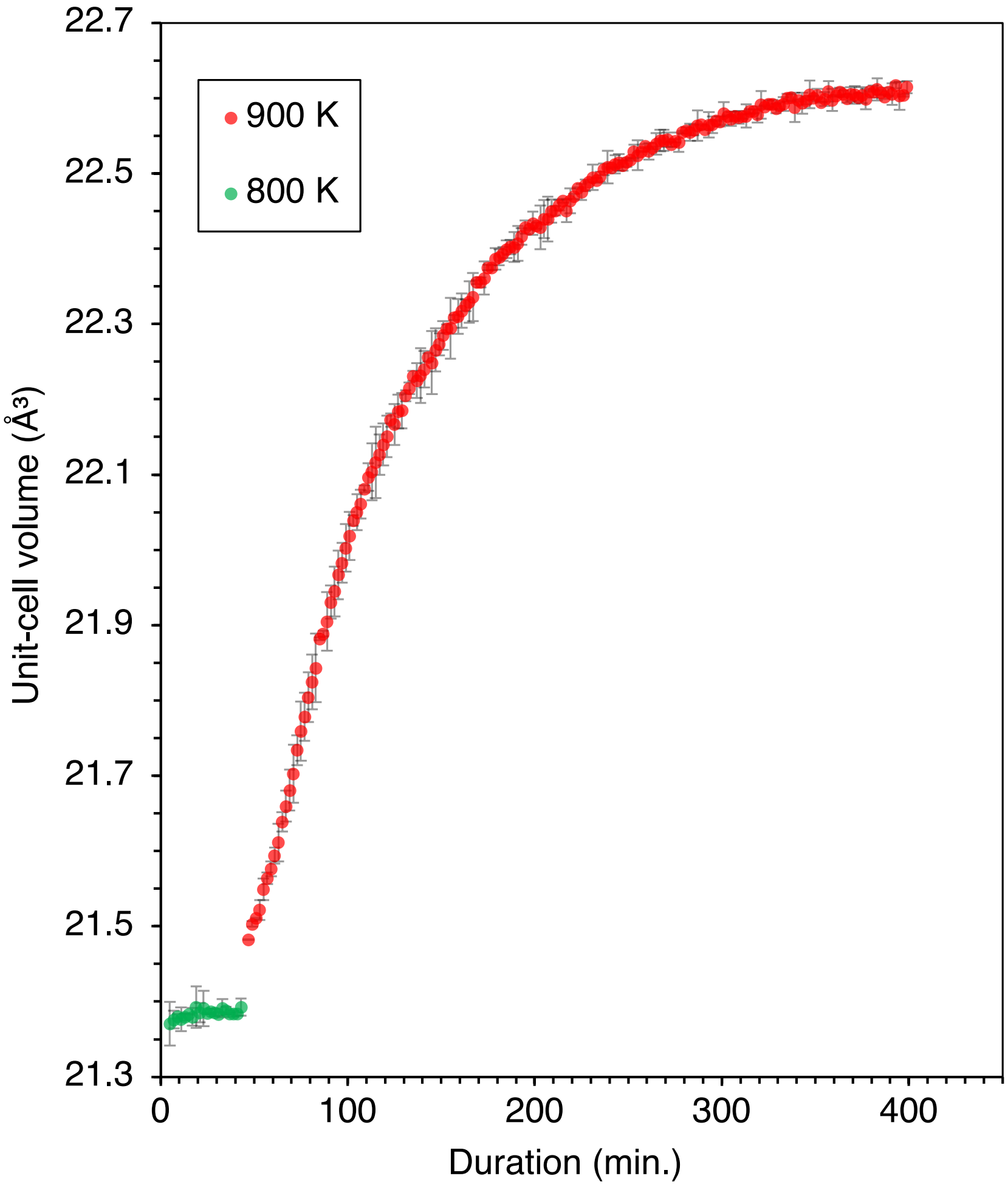

Figure A1: Hydrogenation of hcp-Fe in the supplementary experiment (M4614). First, we heated the sample to 800 K and monitored the unit-cell volume for 50 minutes. At the end of holding at 800 K, the unit-cell volume of the sample (hcp-$FeH_x$) is indistinguishable from that of hcp-Fe because 800 K is just below the hydrogen-releasing temperature. Finally, in the hydrogenation process, the temperature was raised to 900 K. The unit-cell volume of the sample began to expand significantly at 900 K. This observation is reproducible in the experiment used to determine the EoS in this manuscript (M4339). The terminal value of hydrogen content ($x$) in hcp-$FeH_x$ during hydrogenation at 12 GPa and 900 K was derived to be $x$~0.24 using our constructed $v_H(P, T)$ of hcp-$FeH_{0.34(2)}$. No change in unit-cell volume beyond the possible error was observed during the last one hour of hydrogenation.

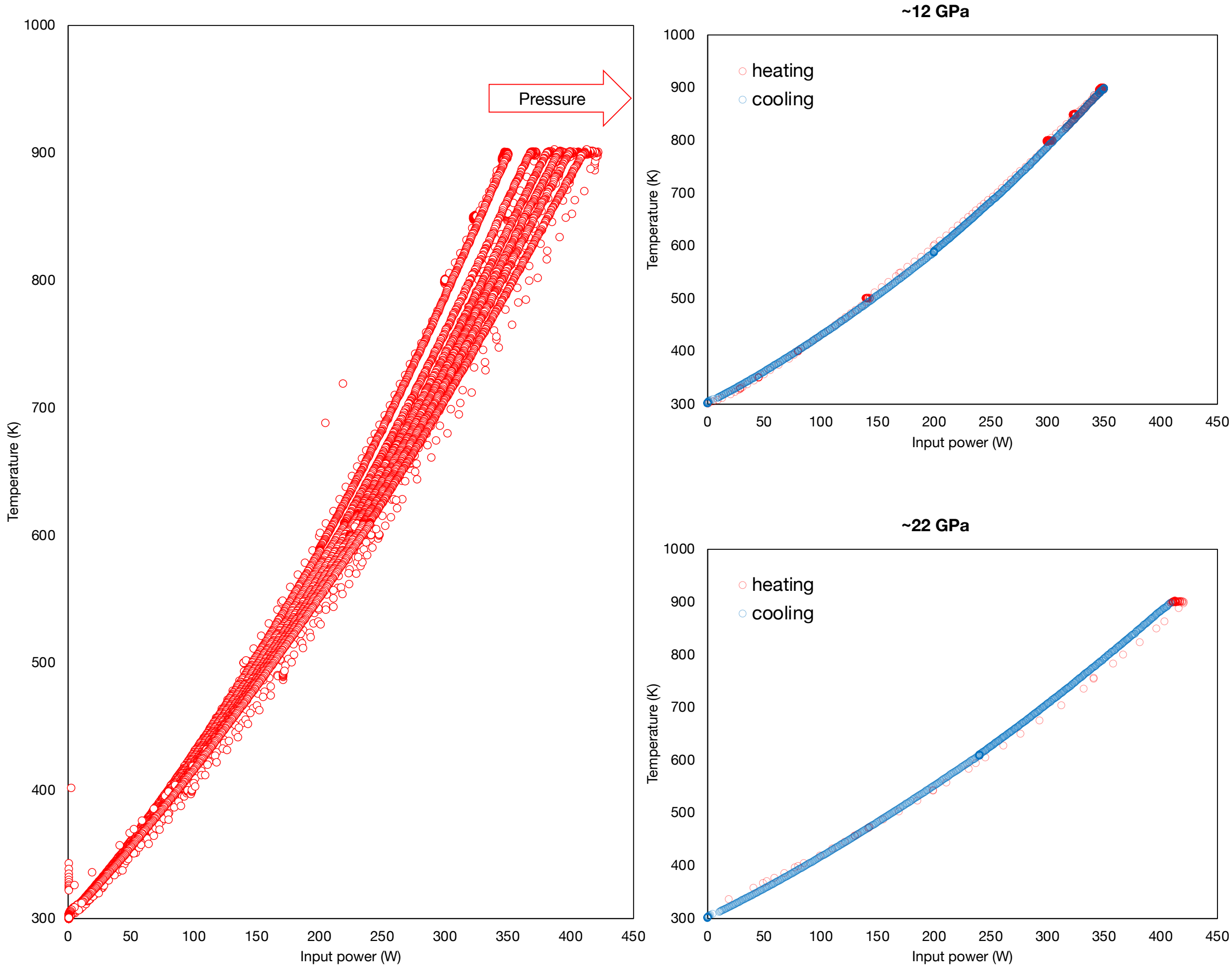

Figure A2: Power–temperature relation of the used cell assembly (M4339). The temperature of the sample is monitored every 10 seconds. Some temperature jumps were detected; however, these points do not show the measured temperature and are thought to be due to temporary electromagnetic noise on the thermocouple. The relationship between temperature and heater resistance shows a stable curve. The whole dataset is plotted in the left figure. The right figure shows the power–temperature relation at ~12 GPa (lowest pressure path) and at ~22 GPa (highest pressure path). The required power to increase the temperature to a certain amount becomes less efficient with increasing pressure. This relation shows almost no hysteresis in the power-temperature relation at any pressure.

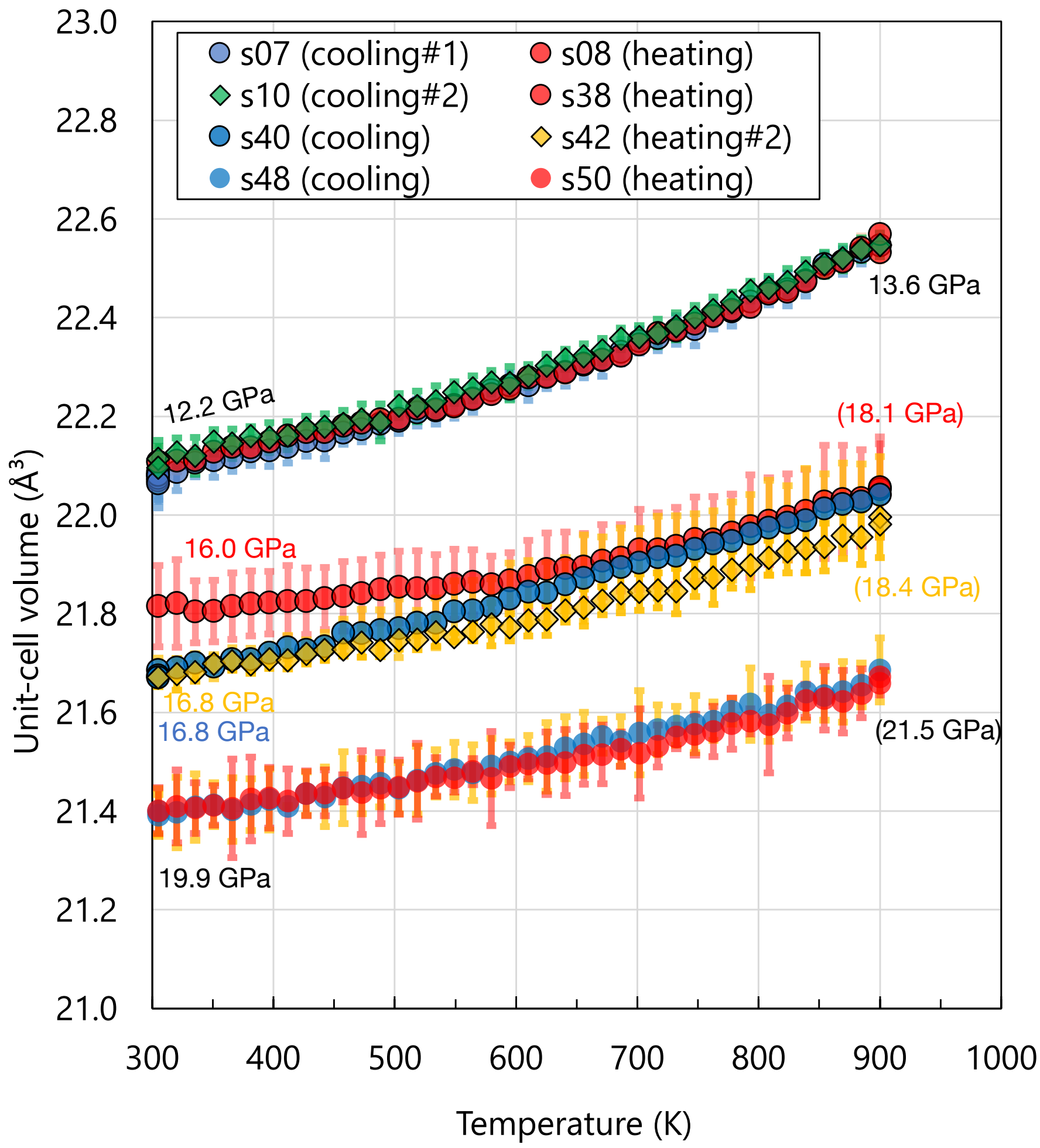

Figure A3: The obtained unit-cell volume of heating and cooling paths. Pressure was determined at 900 K and 300 K and interpolate pressures assuming that it increases proportionally to the temperature increment, which is supported by the preliminary experiments using the same assemblies. In this follow-up experiment, we examined three pressure regions: (i) ~12–14 GPa (s07, s08 and s10), (ii) ~16–18 GPa (s38, s40 and s42) and (iii) ~20–22 GPa (s48 and s50). In the pressure region (ii), there seemed a small inconsistency between heating and cooling processes. This can be explained by the pressure difference in heating and cooling paths, and the volume difference can be explained by the change in pressure, not a change in hydrogen content. Note that the temperature in the cell is calculated from the power–temperature relation because of the thermocouple failure during hydrogenation.

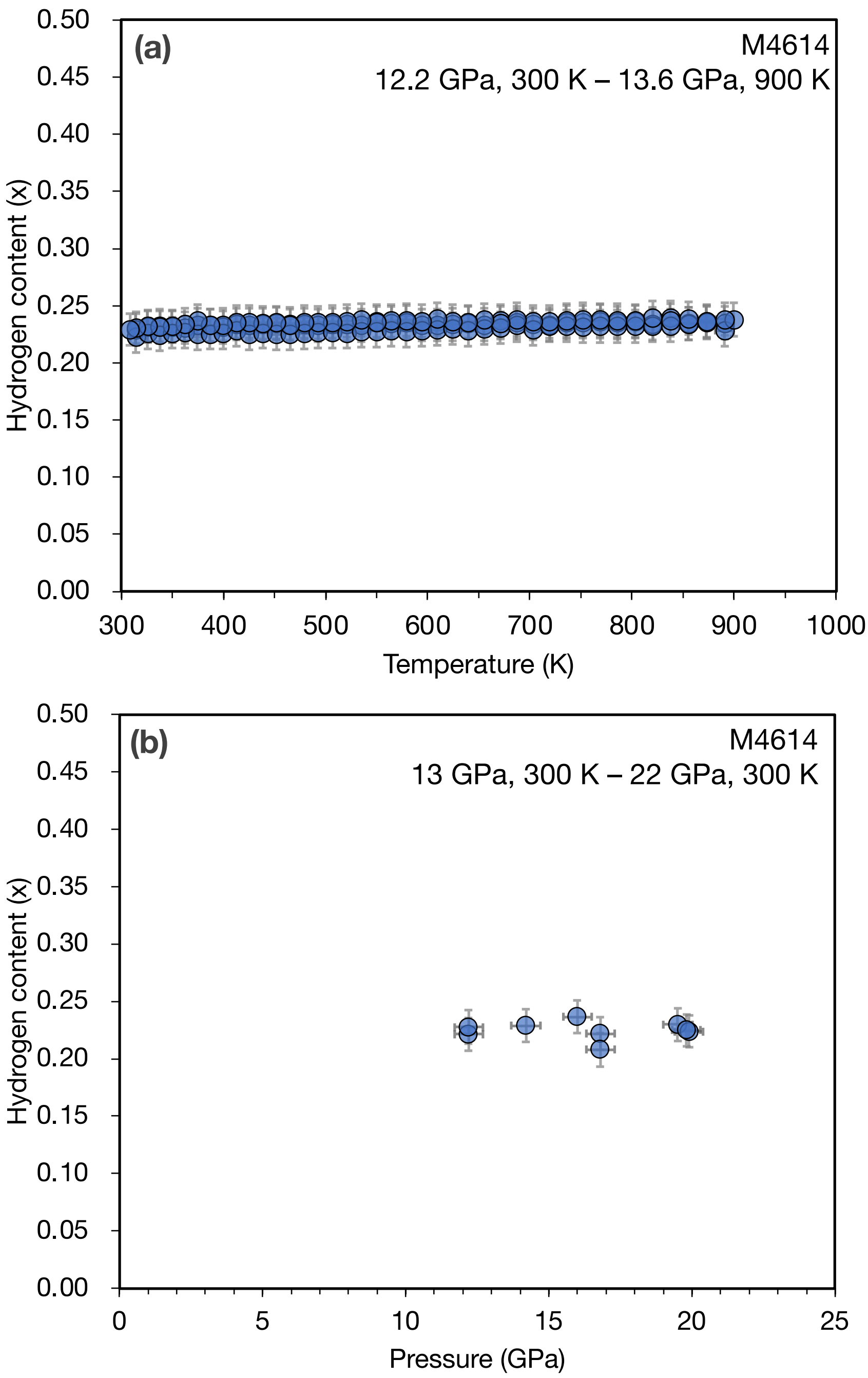

Figure A4: Hydrogen content in hcp-FeH$_x$ in the supplementary experiment (M4614). (a) Hydrogen content in hcp-FeH$_x$ during heating and cooling paths at 12–14 GPa. (b) Hydrogen content in hcp-FeH$_x$ during compression and after heating at room temperature. To calculate hydrogen content in hcp-FeH$_x$, we used our modeled *PT*-dependent $v_H(P, T)$ of hcp-FeH$_{0.34(2)}$. Using the $v_H(P, T)$ obtained in this study, it was shown that the amount of hydrogen in hcp-FeH$_x$ remains constant throughout compression and during heating and cooling processes. As the EoS for hcp-Fe, we used that of Yamazaki et al. (2012).

**Supporting Information**

The supplemental data and analysis related to the manuscript is attached to Supporting Information (Figs.S1–S7, Table S1 and Text S1).

**Data Availability Statement**

Open Research Section: X-ray diffraction data were analyzed using PDIndexer software [Seto et al., 2010; 2025]. The refined lattice constants are listed in Mori (2025).

https://doi.org/10.5281/zenodo.15063628

**Acknowledgments**

Y.M. thank H. Kobayashi (UTokyo) and T. Yagi (UTokyo) for the constructive feedback.

We thank S. Nobusada (UTokyo) and H. Kitagawa (Waseda Univ.) for supporting experiments. This work was supported by a Grant-in-Aid for Encouragement of Research Fellows (22J21462) to Y.M., a JSPS Grant in-Aid for Scientific Research (23H00140) to H.K., and JSPS Grant-in-Aid for Scientific Research (18H05224, 23K13202) to S.K.

Preliminary works for the establishment of the system and the *PVT* experiments were conducted at BL04B1, SPring-8 (project numbers: 2023B0314, 2024A0314, and 2024B0314 to Y.M., 2025B1678 to M.T., and 2022A1319, 2022A2067, 2022B2118, 2023A1250 to S. K.) and NE7, PF-AR, KEK (project numbers: 2023G668 to H.K.).

**Author Contribution**

Y.M., M.T., and H.K. conducted X-ray diffraction experiments at BL04B1, SPring-8, Japan. K.S., N.T., and Y.H assisted the experiments and provided the resources. Y.M. analyzed the data and wrote the draft, and M.T. H.K. and K.A. polished it. All authors reviewed and contributed to manuscript preparation. This study was designed by Y.M.

## References

1. Alfè, D., Gillan, M., & Price, G. D. (2002). Composition and temperature of the Earth's core constrained by combining ab initio calculations and seismic data. *Earth Planet. Sci. Lett.*, 195 (1-2), 91–98.
2. Anderson, O. L., Dubrovinsky, L., Saxena, S. K., & LeBihan, T. (2001). Experimental vibrational Grüneisen ratio values for ω-iron up to 330 GPa at 300 K. *Geophys. Res. Lett.*, 28 (2), 399–402.
3. Antonov, V., Cornell, K., Fedotov, V., Kolesnikov, A., Ponyatovsky, E., Shiryaev, V., & Wipf, H. (1998). Neutron diffraction investigation of the dhcp and hcp iron hydrides and deuterides. *J. Alloys Compd.*, 264 (1-2), 214–222.
4. Badding, J. V., Hemley, R. J., & Mao, H. K. (1991). High-pressure chemistry of hydrogen in metals: In situ study of iron hydride. *Science*, 253(5018), 421-424.
5. Bi, H., Sun, D., Sun, N., Mao, Z., Dai, M., & Hemingway, D. (2025). Seismic detection of a 600-km solid inner core in Mars. *Nature*, 645(8079), 67-72.
6. Caracas, R. (2015). The influence of hydrogen on the seismic properties of solid iron. *Geophys. Res. Lett.*, 42 (10), 3780–3785.
7. Chudinovskikh, L., & Boehler, R. (2007). Eutectic melting in the system Fe–S to 44 GPa. *Earth Planet. Sci. Lett.*, 257(1-2), 97-103.
8. Da Silva, J. R. G., & Mclellan, R. B. (1976). The solubility of hydrogen in super-pure-iron single crystals. *J. Less-Common Met.*, 50(1), 1-5.
9. Dorogokupets, P. I., Dymshits, A. M., Litasov, K. D., & Sokolova, T. S. (2017). Thermodynamics and equations of state of iron to 350 GPa and 6000 K. *Sci. Rep.*, 7(1), 41863.
10. Fei, Y., Murphy, C., Shibazaki, Y., Shahar, A., & Huang, H. (2016). Thermal equation of state of hcp-iron: Constraint on the density deficit of Earth's solid inner core. *Geophys. Res. Lett.*, 43 (13), 6837–6843.
11. Fischer, R. A., Campbell, A. J., Reaman, D. M., Miller, N. A., Heinz, D. L., Dera, P., & Prakapenka, V. B. (2013). Phase relations in the Fe–FeSi system at high pressures and temperatures. *Earth Planet. Sci. Lett.*, 373, 54-64.
12. Fu, S., Chariton, S., Prakapenka, V. B., Chizmeshya, A., & Shim, S. H. (2022). Hydrogen solubility in FeSi alloy phases at high pressures and temperatures. *Amer. Mineral.* 107(12), 2307-2314.

13. Fukai, Y., & Suzuki, T. (1986). Iron-water reaction under high pressure and its implication in the evolution of the Earth. J. *Geophys. Res.: Solid Earth*, 91(B9), 9222-9230.
14. Fukai, Y. (2006). The metal-hydrogen system: basic bulk properties (Vol. 21). Springer Science & Business Media.
15. Genova, A., Goossens, S., Mazarico, E., Lemoine, F. G., Neumann, G. A., Kuang, W., Sabaca, T. J., Hauck II, S. A., Smith, D. E., Solomon, S. C. & Zuber, M. T. (2019). Geodetic evidence that Mercury has a solid inner core. *Geophys. Res. Lett.*, 46(7), 3625-3633.
16. Gomi, H., Fei, Y., & Yoshino, T. (2018). The effects of ferromagnetism and interstitial hydrogen on the equation of states of hcp and dhcp FeHx: Implications for the Earth's inner core age. *Amer. Mineral.*, 103(8), 1271-1281.
17. Hirose, K., Wood, B., & Vočadlo, L. (2021). Light elements in the Earth’s core. *Nat. Rev. Earth Environ*., 2 (9), 645–658.
18. Ikuta, D., Ohtani, E., Sano-Furukawa, A., Shibazaki, Y., Terasaki, H., Yuan, L., & Hattori, T. (2019). Interstitial hydrogen atoms in face-centered cubic iron in the Earth’s core. *Sci. Rep*., 9 (1), 7108.
19. Khan A, Huang D, Durán C, Sossi PA, Giardini D, Murakami M. (2023). Evidence for a liquid silicate layer atop the Martian core. *Nature* 622(7984):718–23
20. Kakizawa, S., Shito, C., Mori, Y., Saitoh, H., Aoki, K., & Kagi, H. (2021). Revised α/ε′–γ phase boundaries for the Fe–H system. *Solid State Commun.*, 340, 114542.
21. Kanzaki, M. (2010). Crystal structure of a new high-pressure polymorph of topaz-oh. *Amer. Mineral.*, 95 (8-9), 1349–1352.
22. Kato, C., Umemoto, K., Ohta, K., Tagawa, S., Hirose, K., & Ohishi, Y. (2020). Stability of fcc phase FeH to 137 GPa. *Amer. Mineral.*, 105(6), 917-921.
23. Katsura, T., Funakoshi, K.-I., Kubo, A., Nishiyama, N., Tange, Y., Sueda, Y.-I., Kubo, T., & Utsumi, W. (2004). A large-volume high-pressure and high-temperature apparatus for in situ X-ray observation,‘SPEED-Mk. II’. *Phys. Earth Planet. Inter.*, 143, 497–506.
24. Kuwayama, Y., Morard, G., Nakajima, Y., Hirose, K., Baron, A. Q., Kawaguchi, S. I., Tsuchiya, T., Ishikawa, D. & Ohishi, Y. (2020). Equation of state of liquid iron under extreme conditions. *Phys. Rev. Lett.*, 124(16), 165701.
25. Machida, A., Saitoh, H., Hattori, T., Sano-Furukawa, A., Funakoshi, K.-I., Sato, T., Orimo, S.-I., & Aoki, K. (2019). Hexagonal close-packed iron hydride behind the conventional phase diagram. *Sci. Rep.*, 9 (1), 1–9.

26. Machida, A., Saitoh, H., Sugimoto, H., Hattori, T., Sano-Furukawa, A., Endo, N., Katayama, Y., Iizuka, R., Sato, T., Matsuno, M., Orimo, S.-I., & Aoki, K. (2014). Site occupancy of interstitial deuterium atoms in face-centred cubic iron. *Nat. Commun.*, 5 (1), 5063.
27. Margot, J. L., Peale, S. J., Jurgens, R. F., Slade, M. A., & Holin, I. V. (2007). Large longitude libration of Mercury reveals a molten core. *Science*, 316(5825), 710-714.
28. Mashino, I., Miozzi, F., Hirose, K., Morard, G., & Sinmyo, R. (2019). Melting experiments on the Fe–C binary system up to 255 GPa: Constraints on the carbon content in the Earth's core. *Earth Planet. Sci. Lett.*, 515, 135-144.
29. Matsui, M., Higo, Y., Okamoto, Y., Irifune, T., & Funakoshi, K.-I. (2012). Simultaneous sound velocity and density measurements of NaCl at high temperatures and pressures: Application as a primary pressure standard. *Amer. Mineral.*, 97 (10), 1670–1675.
30. Matsuoka, T., Muraoka, S., Ishikawa, T., Niwa, K., Ohta, K., Hirao, N., Kawaguchi, S., Ohishi, Y., Shimizu, K., & Sasaki, S. (2019). Hydrogen-storing salt NaCl ($H_2$) synthesized at high pressure and high temperature. *J. Phys. Chem. C*, 123 (41), 25074–25080.
31. Miozzi, F., Matas, J., Guignot, N., Badro, J., Siebert, J., & Fiquet, G. (2020). A new reference for the thermal equation of state of iron. *Minerals*, 10 (2), 100.
32. Mita, S., Tagawa, S., Hirose, K., & Ikuta, N. (2025). Fe-FeH eutectic melting curve and the estimates of Earth's core temperature and composition. *J. Geophys. Res.: Solid Earth*, 130(1), e2024JB029283.
33. Mori, Y., Kagi, H., Kakizawa, S., Komatsu, K., Shito, C., Iizuka–Oku, R., Aoki, K., Hattori, T Sano-Furukawa, A., Funakoshi, K.-I. & Saitoh, H. (2021). Neutron diffraction study of hydrogen site occupancy in $Fe_{0.95}Si_{0.05}$ at 14.7 GPa and 800 K. *J. Mineral. Petrol, Sci.*, 116(6), 309-313.
34. Mori, Y. (2025). Thermal equation of state of iron hydride with hcp structure – its implications for density deficit in the core [Dataset]. *Zenodo*, https://doi.org/10.5281/zenodo.15063628

35. Mori, Y., Aoki, K., Takano, M., Kagi, H., Park, I., Wang, Z., Kim, D.Y., Tsujino, N., Kakizawa, S. & Higo, Y. (2025). Invar behavior and negative thermal expansion linked to magnetic transition in dhcp iron hydride under high pressure conditions. *arXiv*, 2501, 08937.
36. Mori, Y., Kagi, H., Aoki, K., Takano, M., Kakizawa, S., Sano-Furukawa, A., & Funakoshi, K.-I. (2024). Hydrogenation of silicon-bearing hexagonal close-packed iron and its implications for density deficits in the inner core. *Earth Planet. Sci. Lett.*, 634, 118673.
37. Nylén, J., Sato, T., Soignard, E., Yarger, J. L., Stoyanov, E., & Häussermann, U. (2009). Thermal decomposition of ammonia borane at high pressures. *J. Chem. Phys.*, 131 (10).
38. Oka, K., Ikuta, N., Tagawa, S., Hirose, K., & Ohishi, Y. (2022). Melting experiments on Fe-O-H and Fe-H: Evidence for eutectic melting in Fe-FeH and implications for hydrogen in the core. *Geophys. Res. Lett*, 49(17), e2022GL099420.
39. Okuchi, T. (1997). Hydrogen partitioning into molten iron at high pressure: implications for Earth's core. *Science*, 278 (5344), 1781–1784.
40. Ono, S. (2015). Relationship between structural variation and spin transition of iron under high pressures and high temperatures. *Solid State Commun.*, 203, 1–4.
41. Pépin, C. M., Geneste, G., Dewaele, A., Mezouar, M., & Loubeyre, P. (2017). Synthesis of $FeH_5$: A layered structure with atomic hydrogen slabs. *Science*, 357(6349), 382-385.
42. Plattner, A. M., & Johnson, C. L. (2021). Mercury's Northern Rise core-field magnetic anomaly. *Geophys. Res. Lett.*, 48(17), e2021GL094695.
43. Sakai, F., & Hirose, K. (2026). Melting phase relations in Fe-FeS under Martian core pressures: Crystallization of $Fe_{12}S_7$ at the inner core? *Earth Planet. Sci. Lett.*, 675, 119785.
44. Sakamaki, K., Takahashi, E., Nakajima, Y., Nishihara, Y., Funakoshi, K., Suzuki, T., & Fukai, Y. (2009). Melting phase relation of FeHx up to 20 GPa: Implication for the temperature of the Earth's core. *Phys. Earth Planet. Inter.*, 174(1-4), 192-201.
45. Samuel H, Drilleau M, Rivoldini A, Xu Z, Huang Q, et al. (2023). Geophysical evidence for an enriched molten silicate layer above Mars's core. *Nature* 622(7984):712–17
46. Sano-Furukawa, A., Kakizawa, S., Shito, C., Hattori, T., Machida, S., Abe, J., Funakoshi, K.-I. & Kagi, H. (2021). High-pressure and high-temperature neutron-diffraction experiments using Kawai-type multi-anvil assemblies. *High Pressure Res.*, 41(1), 65-74.

47. Saxena, S. K., & Dubrovinsky, L. S. (2000). Iron phases at high pressures and temperatures: Phase transition and melting. *Amer. Mineral.*, 85(2), 372-375.
48. Seto, Y.; Nishio-Hamane, D.; Nagai, T.; Sata, N. Development of a Software Suite on X-ray Diffraction Experiments. *Rev. High Press. Sci. Technol.* 2010, 20 (3), 269−276.
49. Seto, Y. (2025) PDIndexer (version 4.450) [Software]. https://github.com/seto77/PDIndexer/.
50. Sha, X., & Cohen, R. (2010). First-principles thermal equation of state and thermoelasticity of hcp Fe at high pressures. *Phys. Rev. B*, 81 (9), 094105.
51. Shahar, A., Young, E. D., Hirose, K., & Yokoo, S. (2026). The Compositions of Planetary Cores. *Annu. Rev. Earth Planet. Sci.,* 54.
52. Shen, G., Sturhahn, W., Alph, E., Zhao, J., Tollenner, T., Prakapenka, V., Meng, Y. & Mao, H.-R. (2004). Phonon density of states in iron at high pressures and high temperatures. *Phys. Chem. Miner.*, 31, 353–359.
53. Sinmyo, R., Hirose, K., & Ohishi, Y. (2019). Melting curve of iron to 290 GPa determined in a resistance-heated diamond-anvil cell. *Earth Planet Sci. Lett.*, 510, 45-52.
54. Skorodumova, N., Ahuja, R., & Johansson, B. (2004). Influence of hydrogen on the stability of iron phases under pressure. *Geophys. Res. Lett.*, 31 (8).
55. Stähler SC, Khan A, Banerdt WB, Lognonné P, Giardini D, et al. (2021). Seismic detection of the Martian core. *Science* 373(6553):443–48
56. Stewart, A. J., Schmidt, M. W., Van Westrenen, W., & Liebske, C. (2007). Mars: A new core-crystallization regime. *Science*, 316(5829), 1323-1325.
57. Stoutenburg, E. R., Caracas, R., & Campbell, A. J. (2026). Immiscibility between hydrogen and molten iron in planetary cores. *Earth Planet. Sci. Lett.*, 678, 119851.
58. Tagawa, S., Sakamoto, N., Hirose, K., Yokoo, S., Hernlund, J., Ohishi, Y., & Yurimoto, H. (2021). Experimental evidence for hydrogen incorporation into Earth’s core. *Nat. Commun.*, 12 (1), 2588.
59. Tagawa, S., Gomi, H., Hirose, K., & Ohishi, Y. (2022a). High-temperature equation of state of FeH: Implications for hydrogen in Earth's inner core. *Geophys. Res. Lett.*, 49(5), e2021GL096260.
60. Tagawa, S., Helffrich, G., Hirose, K., & Ohishi, Y. (2022b). High-pressure melting curve of FeH: Implications for eutectic melting between Fe and non-magnetic FeH. *J. Geophys. Res.: Solid Earth*, 127 (6), e2022JB024365.

61. Terasaki, H., Shibazaki, Y., Sakamaki, T., Tateyama, R., Ohtani, E., Funakoshi, K. I., & Higo, Y. (2011). Hydrogenation of FeSi under high pressure. *Amer. Mineral.*, 96(1), 93-99.
62. Takano, M., Kagi, H., Mori, Y., Aoki, K., Kakizawa, S., Sano-Furukawa, A., Iizuka-Oku, R. & Tsuchiya, T. (2024). Low reactivity of stoichiometric FeS with hydrogen at high-pressure and high-temperature conditions. *J. Mineral. Petrol. Sci.*, 119(1), 240122.
63. Takano, M., Kagi, H., Mori, Y., Kobayashi, H., Aoki, K., Takahashi, Y., Ijichi, Y., Kakizawa, S., Tsujino, N., Higo, Y. & Tsuchiya, T. (2026). Negligible Hydrogen Solubility in FeS under High-Pressure and High-Temperature Conditions: Implications for Hydrogen Storage in Planetary Cores. *ACS Earth Space Chem.* (in press)
64. Tateno, S., Hirose, K., Ohishi, Y., & Tatsumi, Y. (2010). The structure of iron in Earth's inner core. *Science*, 330(6002), 359–361.
65. Tsujino, N., Nishihara, Y., Nakajima, Y., Takahashi, E., Funakoshi, K. I., & Higo, Y. (2013). Equation of state of $\gamma$-Fe: Reference density for planetary cores. *Earth Planet. Sci. Lett.*, 375, 244-253.
66. Uchida, T., Wang, Y., Rivers, M. L., & Sutton, S. R. (2001). Stability field and thermal equation of state of ε-iron determined by synchrotron X-ray diffraction in a multianvil apparatus. *J. Geophys. Res.: Solid Earth*, 106 (B10), 21799–21810.
67. Yagi, T., & Hishinuma, T. (1995). Iron hydride formed by the reaction of iron, silicate, and water: Implications for the light element of the Earth's core. *Geophys. Res.Lett.,* 22(14), 1933-1936.
68. Yamazaki, D., Ito, E., Yoshino, T., Yoneda, A., Guo, X., Zhang, B., Sun, W., Shimojuku, A., Tsujino, N., Kunimoto, T., Higo, Y., & Funakoshi, K.-I. (2012). P-V-T equation of state for ε-iron up to 80 GPa and 1900 K using the Kawai-type high pressure apparatus equipped with sintered diamond anvils. *Geophys. Res. Lett.*, 39 (20).
69. Yokoo, S., Hirose, K., Tagawa, S., Morard, G., & Ohishi, Y. (2022). Stratification in planetary cores by liquid immiscibility in Fe-S-H. *Nat. Commun.*, 13(1), 644.
70. Yuan, L., & Steinle-Neumann, G. (2023). Hydrogen distribution between the Earth's inner and outer core. *Earth Planet. Sci. Lett.*, 609, 118084.
71. Zhang, Y., Wang, W., Li, Y., & Wu, Z. (2024). Superionic iron hydride shapes ultralow-velocity zones at Earth's core–mantle boundary. *Proc. Natl. Acad. Sci. U.S.A.,* 121(35), e2406386121.

72. Zhang, Z., Wang, W., Liu, J., Zhang, Y., Mitchell, R. N., & Zhang, Z. (2025). Oxygen driving hydrogen into the inner core: Implications for the Earth's core composition. *Geophys. Res. Lett.*, 52(3), e2024GL110315.
73. Zhang, Y., Zhang, S., Kuang, D., & Xiong, C. (2025). Equation of State Parameters of hcp-Fe up to Super-Earth Interior Conditions. *Crystals*, 15 (3), 221.

## Supporting Information

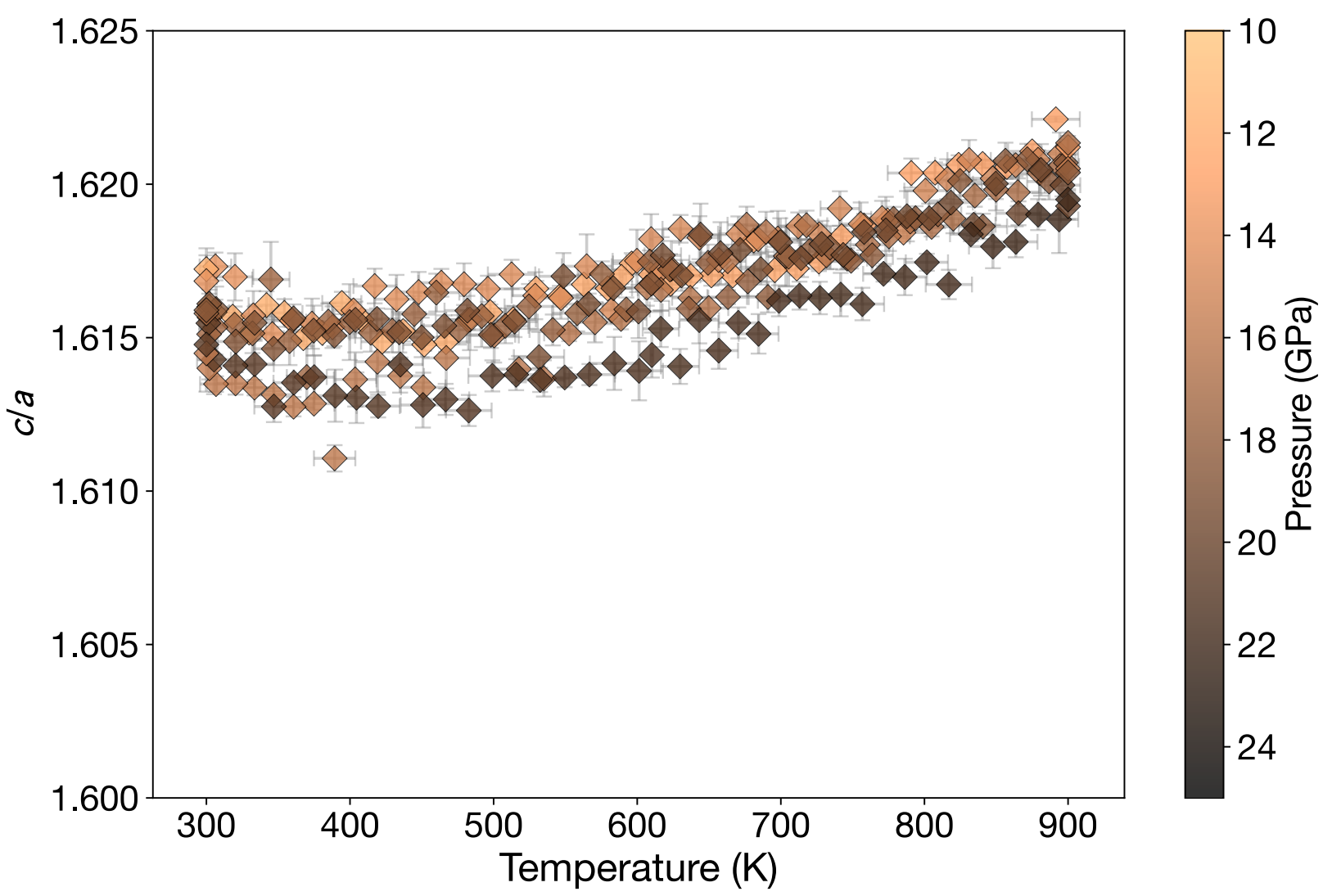

**Figure S1**: Axial ratio, *c/a*, of hcp-$FeH_x$ contoured as a function of pressure.

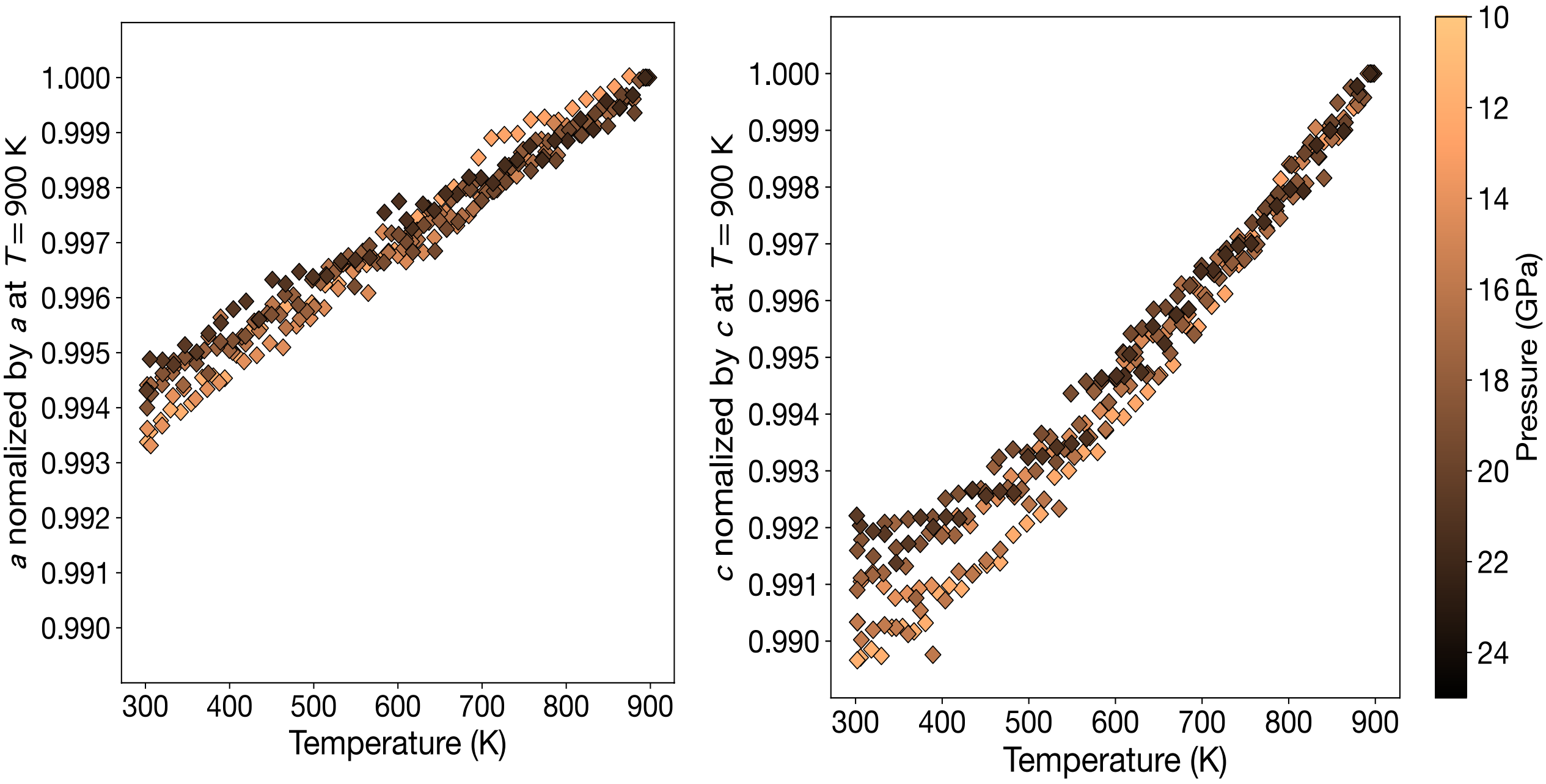

**Figure S2**: Temperature dependence of the lattice constants, *a* and *c*. The volume at the maximum temperature (900 K) in each pass is set to 1, and the volumes during cooling are normalized.

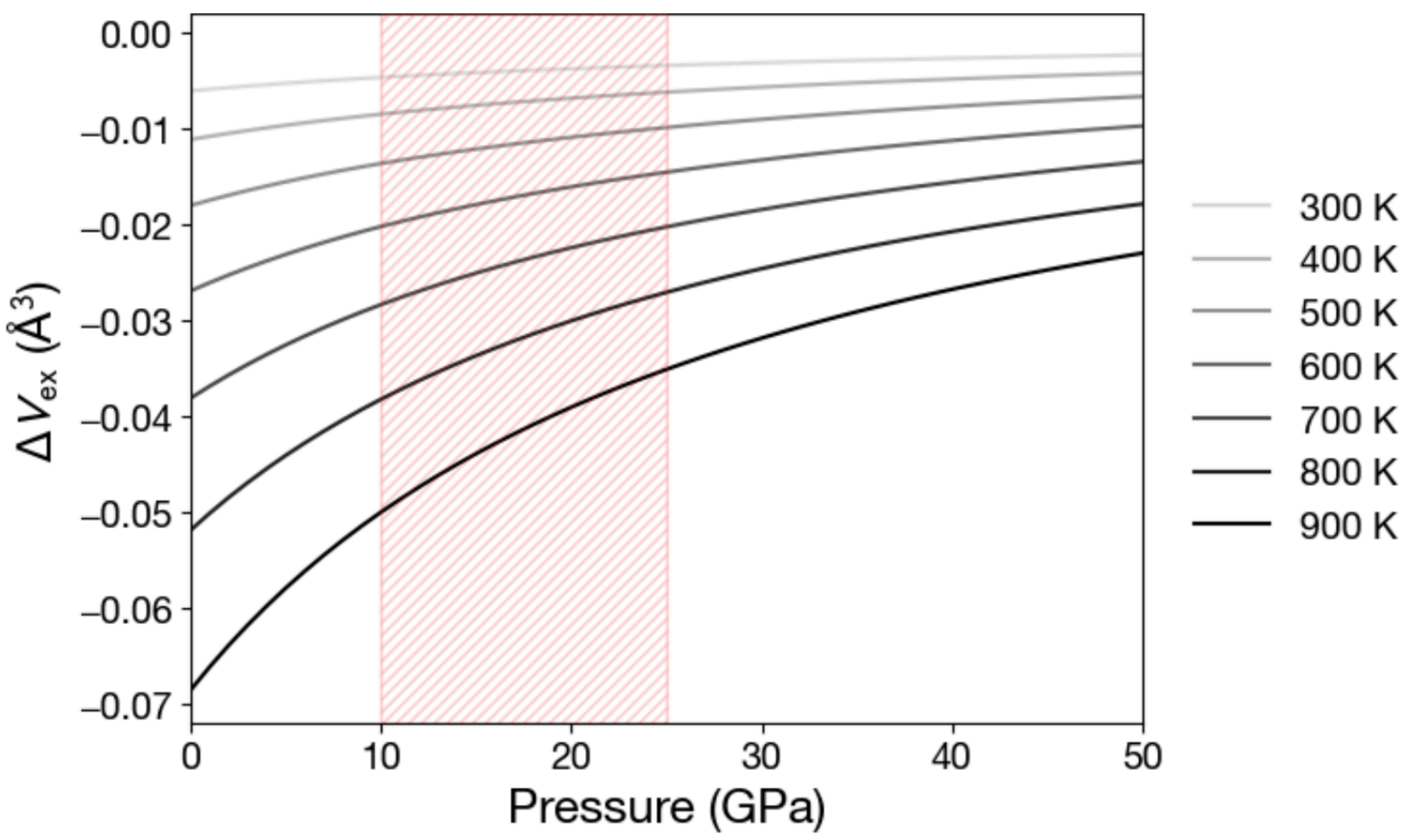

**Figure S3**: The effect of anharmonic and electronic contribution on the unit-cell volume ($\Delta V_{ex}$) of hcp-Fe [Alfè et al., 2001]. In the experimental range (red shaded area), anharmonic and electronic pressure were not significant to determine the EoS of hcp-$FeH_x$. Also, the ratio between $\Delta V_{ex}$ and its unit cell is small, allowing robust estimation of hydrogen-induced density reduction, albeit less effective for estimating $v_H$ at higher temperatures. In hcp-Fe, their effects are predicted to be large compared to bcc Fe [Belonoshko et al., 2009]. Those properties of fcc-Fe are still unknown [Tsujino et al., 2012].

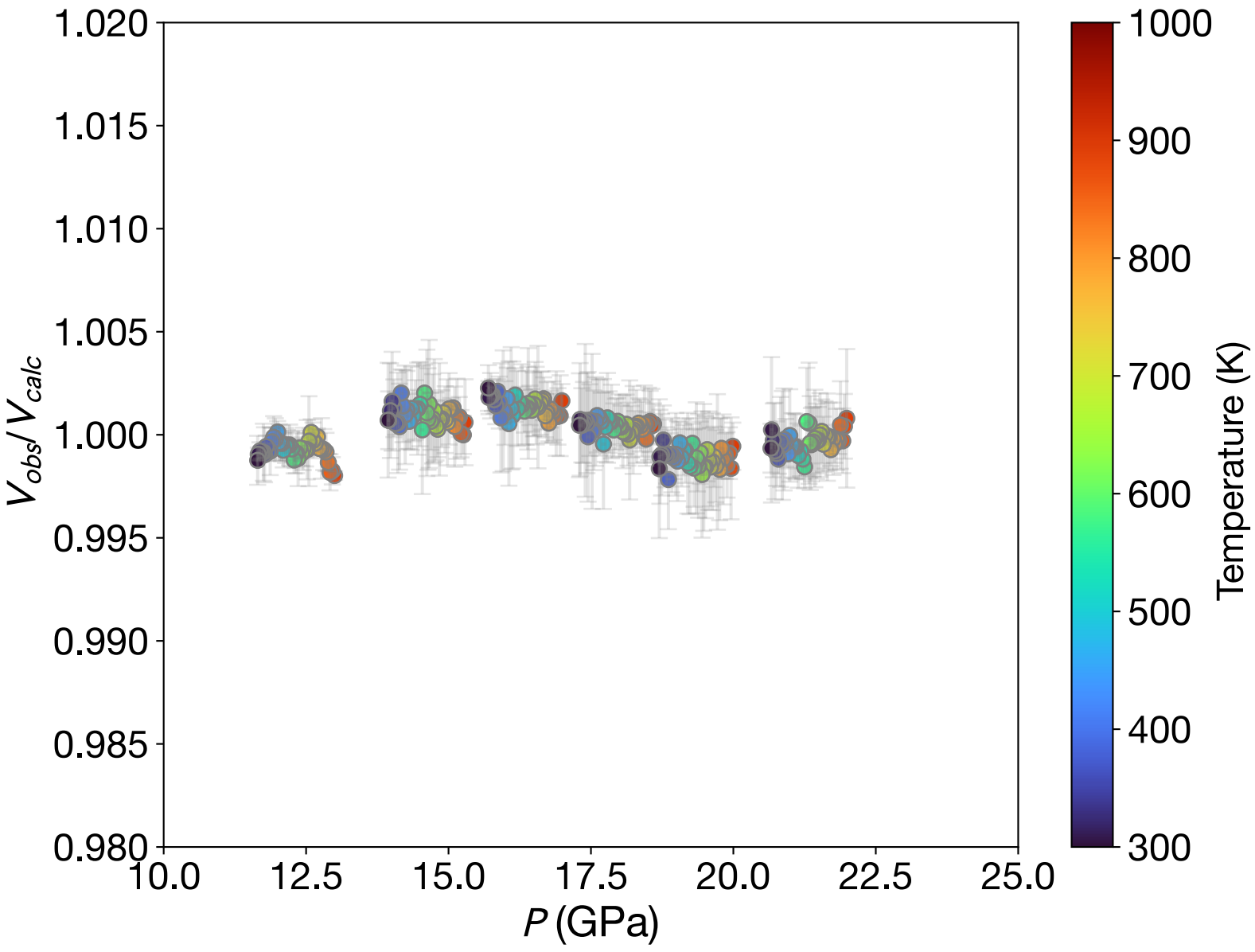

**Figure S4**: Difference between the unit-cell volume by EoS ($V_{calc}$) and the observed unit-cell volume ($V_{obs}$). Here, we fitted the $PVT$ data of hcp-$FeH_x$ to the third-order Birch-Murnaghan EoS with MGD model.

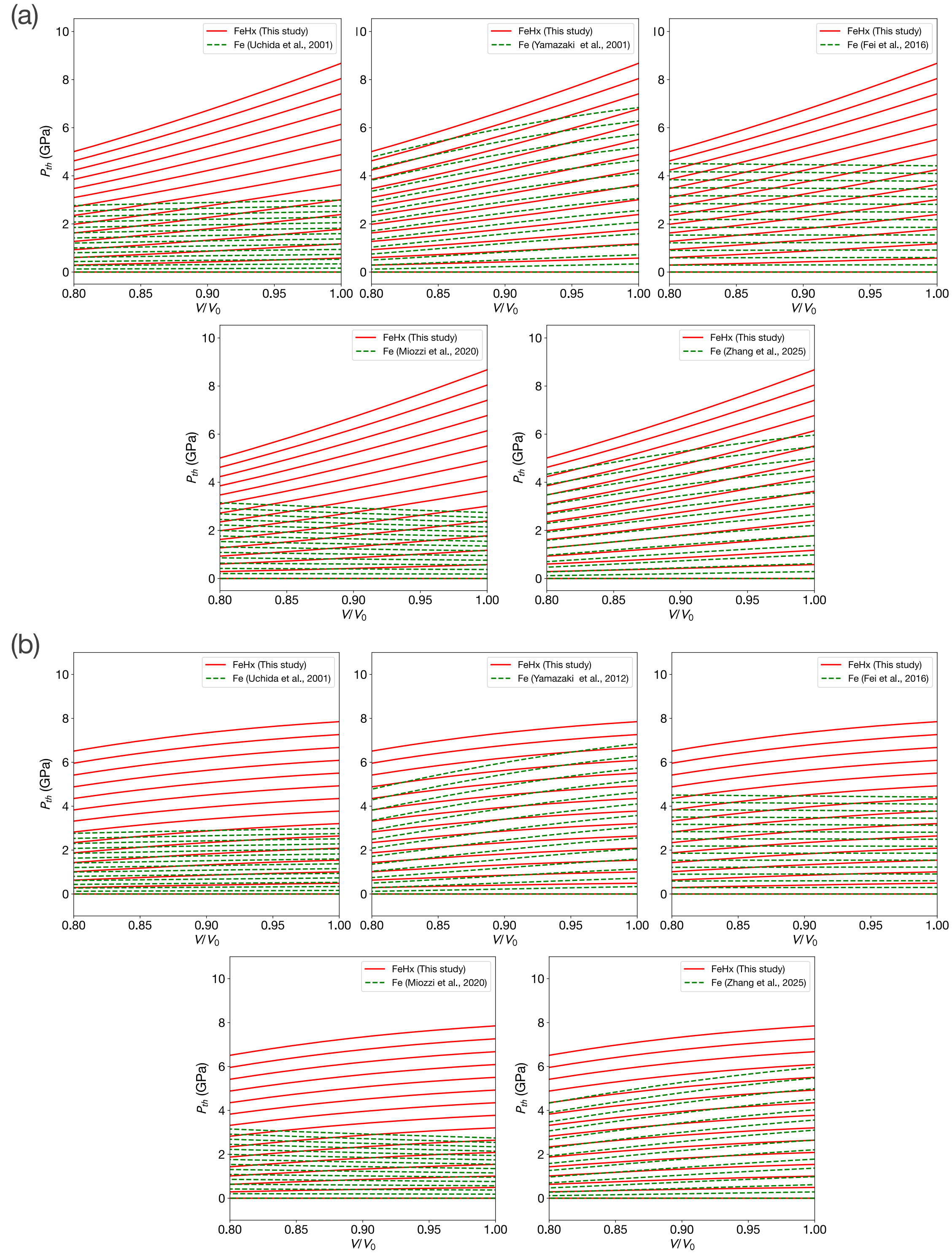

**Figure S5**: Relation between thermal pressure ($P_{th}$) and normalized unit-cell volume ($V$) for hcp-$FeH_x$ (red solid line) and hcp-Fe (green dashed line). Isothermal pressure is calculated every 50 K from 300 K to 1000 K. (a) hcp-$FeH_x$ with $q$ treated as a fitting parameter; (b) hcp-$FeH_x$ with $q$ fixed at 1. Table 1(a) lists the fitted thermoelastic values.

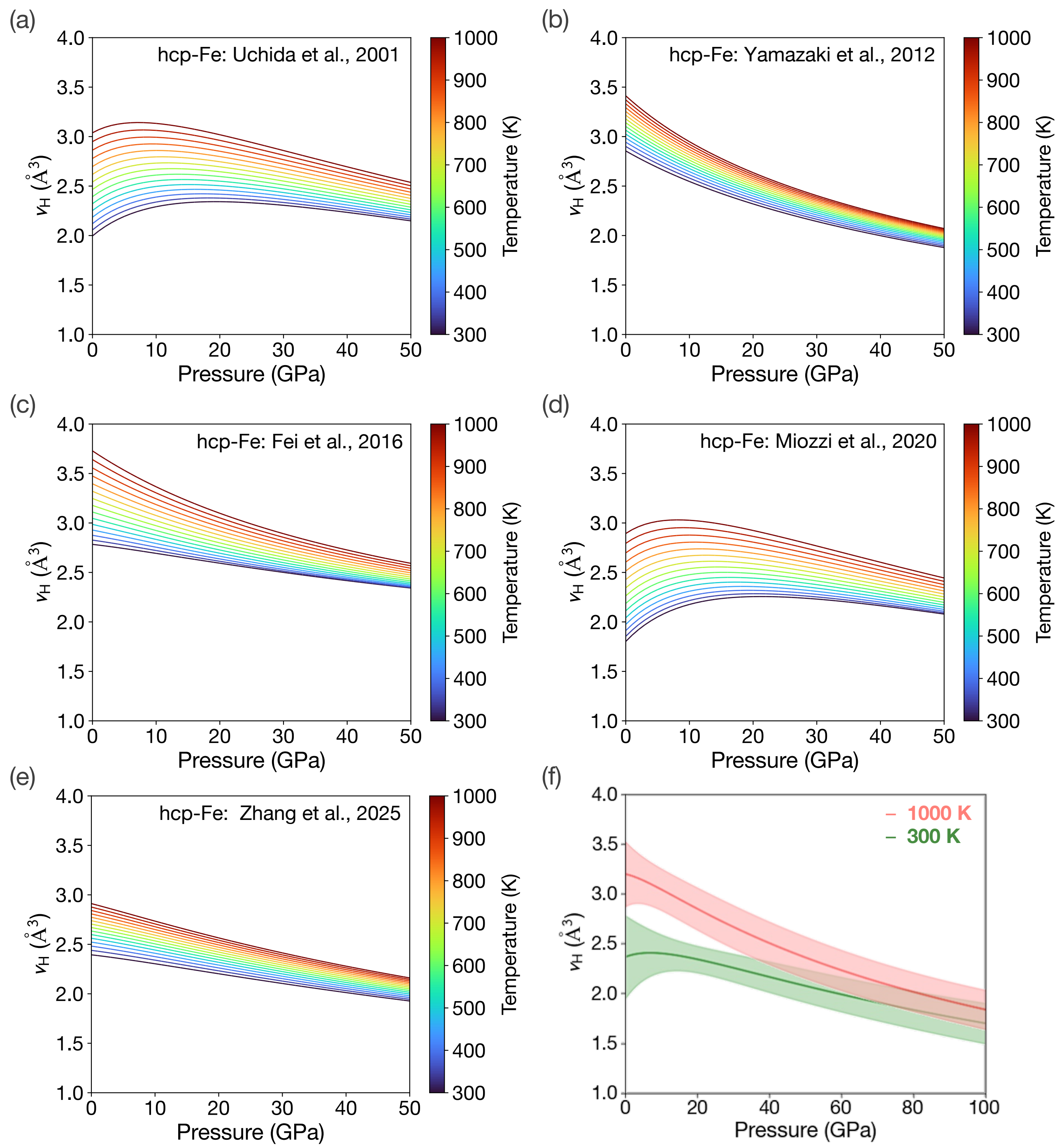

**Figure S6**: *PT* dependence of $v_H$ for hcp-Fe$H_x$ at $P$ = 0–50 GPa and $T$ = 300–1000 K with various EoS for hcp-Fe. Here, we used BM-EoS and MGD approximation, fixing $q$ = 1. Each figure is depicted in the same manner as in Figs. 3 and 4 in the manuscript. As hcp-Fe EoS, we used (a) Uchida et al. (2001), (b) Yamazaki et al. (2012), (c) Fei et al. (2016), (d) Miozzi et al. (2020), and (e) Zhang, Y. et al. (2025).

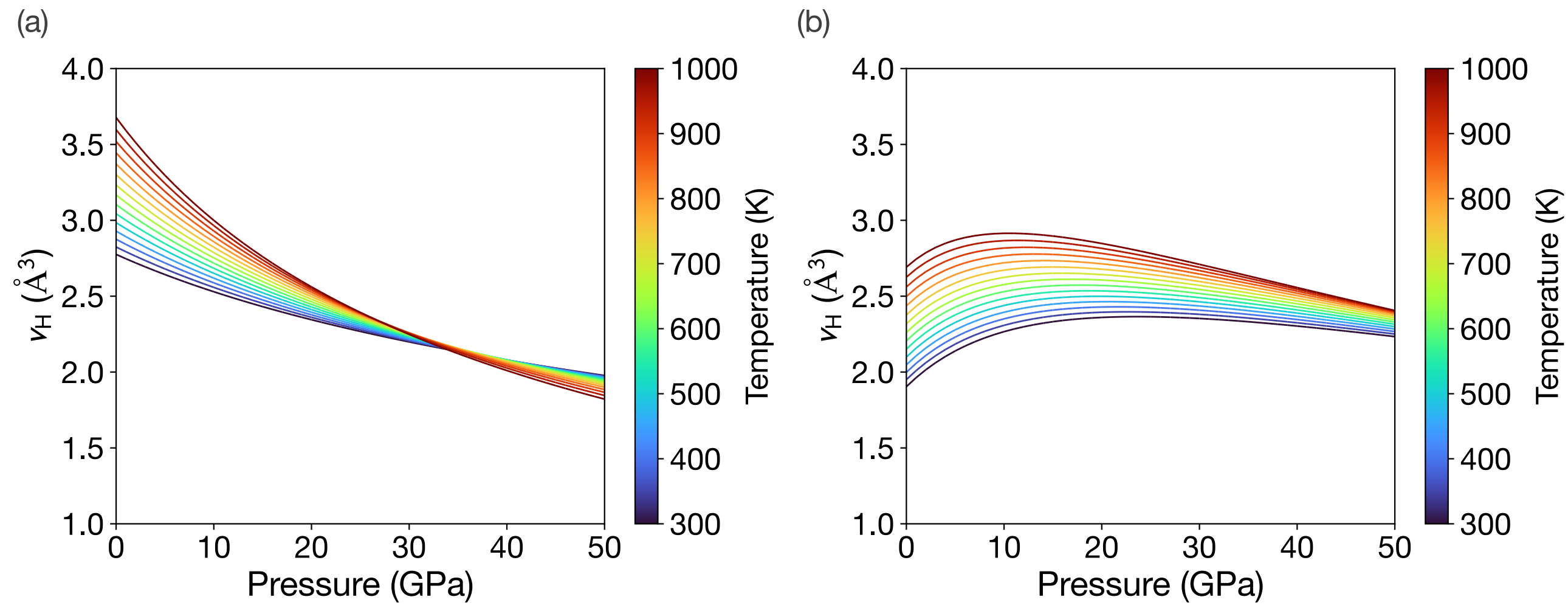

**Figure S7**: Isothermal compression behavior of $v_H$ for hcp-FeH$_x$ at $P$ = 0–50 GPa and $T$ = 300–1000 K by applying Birch-Murnaghan EoS with polynomial thermal expansion. (a) $PT$-dependence of $v_H$ with Yamazaki et al. (2012) and (b) that with Uchida et al. (2001). In both thermoelastic expressions (i.e., MGD and empirical polynomial thermal expansion models), the tendency of the response of $v_H$ to pressure and temperature is similar, but the magnitude of the change varies.

**Table S1**: Elastic parameters of the EoS for hcp-FeH$_x$ with empirical (polynomial) thermal expansion.

| | $V_0$ (Å$^3$) | $K$ (GPa) | $K_T'$ | EoS | $a(10^{-5})$ | $b(10^{-8})$ | $\frac{dK}{dT}_P$ (GPa/K) |
|---|---|---|---|---|---|---|---|
| FeH$_x$ | 23.92(2) | 187(3) | 4.3(fixed) | BM | 6.2(2) | 6.4(11) | -0.077(8) |
| FeH$_x$ | 23.91(2) | 188(3) | 4.3(fixed) | Vinet | 6.2(2) | 6.3(10) | -0.077(8) |

**Text S1: The choice of EoS for hcp-Fe**

We also fitted the *PVT* data of hcp-$FeH_x$ to Vinet equations of state,

$$P(T_0, V) = 3K_0\left(\frac{V_0}{V}\right)^{\frac{2}{3}}\left\{1-\left(\frac{V_0}{V}\right)^{-\frac{1}{3}}\right\}\exp\left[\frac{3}{2}(K_0'-1)\left\{1-\left(\frac{V_0}{V}\right)^{-\frac{1}{3}}\right\}\right]. \quad \text{(S1)}$$

but it has no significant effect on the derived elastic parameters, and isothermal curves of both BM and Vinet are comparable. On the other hand, the thermal expansion model can change the *PT*-dependence of $v_H$. Empirically, the thermal expansion and temperature effect on the bulk modulus are described as follows.

$$V(T, P_0) = V(T_0, P_0)\exp\left\{\int_{T_0}^{T}(a+bT)dT\right\}, \quad \text{(S2)}$$

and

$$K(T, P_0) = K(T_0, P_0) + \frac{\partial K(T, P_0)}{\partial T}(T-T_0). \quad \text{(S3)}$$

The integrand in the RHS of Eq. S2 is the empirical expression of thermal expansion, where *a* and *b* are constants. The standard temperature $T_0$ was set to 300 K. The fitted elastic and thermoelastic parameters are listed in Table S1. If assuming such a polynomial thermal expansion, the temperature dependence of $v_H$ flipped into negative (Fig. S7). Although it is not certain whether this reversal results from pressure extrapolation beyond the experimentally obtained range, the thermal effect on $v_H$ decreases with increasing pressure in both MGD and polynomial thermal expansion models.

The first report of $v_H$ for hcp-$FeH_x$ was by Machida et al. (2019). They calculated vH using the equation of state for Fe by Uchida et al. (2001). As later pointed out by Fei et al. (2016), those pioneering works sometimes exhibit significant scatter. The *PVT* studies by Yamazaki et al. (2012), Fei et al. (2016), and Zhang, Y. et al. (2025) show a similar trend. In contrast, an experimental study by Miozzi et al. (2020) also shows a similar trend to Uchida et al. (2001). That inconsistency can originate from the difference in $V_0$, $K_0$, and $K_0'$. In addition, there are also two trends in Debye temperature: commonly used value in DAC studies is 420 K [Fei et al., 2016; Miozzi et al., 2020], and the fitted value in KMA studies is around 1000 K [Uchida et al., 2001; Yamazaki et al., 2012]. In this study, we do not pursue each trend individually, as this is not the focus of our research, and only mention the general trend for each EoS. Most importantly, regardless of the choice of equations of state (EoS), the hydrogen content within each dataset is internally consistent.

**Additional References in Supporting Information**

74. Alfè, D., Price, G. D. & Gillan, M. J. (2002). Thermodynamics of hexagonal close-packed iron under Earth's core conditions, *Phys. Rev. B,* 64, 045123

75. Belonoshko, A. B., Dorogokupets, P. I., Johansson, B., Saxena, S. K. & Kochi, L. (2008). Ab initio equation of state for the body-centered-cubic phase of iron at high pressure and temperature, *Phys. Rev. B*, 78, 104107